\documentclass[reprint, amsmath, amssymb, aps]{article}
\usepackage[utf8]{inputenc}
\usepackage{amssymb}
\usepackage{fullpage}
\usepackage{amsmath}
\usepackage[english]{babel}
\usepackage{csquotes}
\usepackage{eurosym}
\usepackage{xcolor}
\usepackage{mathrsfs}
\usepackage{graphicx}
\usepackage{caption}
\usepackage{subcaption}
\usepackage{float}
\usepackage{bm}
\usepackage{hyperref}
\usepackage{authblk}
\usepackage[backend=biber, style=numeric, sorting=none, maxbibnames=99]{biblatex}
\newtheorem{theorem}{Theorem}[section]
\author[1]{Anna Büttner \footnote{buettner@pik-potsdam.de}}
\author[1]{Jürgen Kurths}
\author[1]{Frank Hellmann \footnote{hellmann@pik-potsdam.de}}

\affil[1]{Potsdam Institute for Climate Impact Research}
\date{}

\begin{document}
    \title{Ambient Forcing: Sampling Local Perturbations in Constrained Phase Spaces}
    \maketitle
\section*{Abstract}
    Ambient Forcing is a novel method to sample random states from manifolds of differential-algebraic equations (DAE). These states can represent local perturbations of nodes in power systems with loads, which introduces constraints into the system. These states must be \textit{valid} initial conditions to the DAE, meaning that they fulfill the algebraic equations. Additionally, these states should represent perturbations of individual variables in the power grid, such as a perturbation of the voltage at a load.
    These initial states enable the calculation of probabilistic stability measures of power systems with loads, which was not yet possible, but is important as these measures have become a crucial tool in studying power systems. 
    To verify that these perturbations are network local, i.e. that the initial perturbation only targets a single node in the power grid, a new measure, the spreadability, related to the closeness centrality \cite{closeness_centrality}, is presented. The spreadability is evaluated for an ensemble of typical power grids. The ensemble depicts a set of future power grids where consumers, as well as producers, are connected to the grid via inverters.
    For this power grid ensemble, we additionally calculate the basin stability \cite{bs_first} as well as the survivability \cite{surv}, two probabilistic measures which provide statements about asymptotic and transient stability. We also revisit the topological classes, introduced in \cite{topology_algo}, that have been shown to predict the basin stability of grids and explore if they still hold for grids with constraints and voltage dynamics. We find that the degree of the nodes is a better predictor than the topological classes for our ensemble. Finally, ambient forcing is applied to calculate probabilistic stability measures of the IEEE 96 test case \cite{ieee96}.

%%%%%%%%%%%%%%%%%%%%%%%%%%%%%%%%%%%%%%%%%%%%%%%%%%%%%%%%%%%%%%%%%%%%%%%%%%%%%%%%%%%%%%%%%%%%%%%%%

    \section{Introduction}
    Currently, power grids all over the world are reconstructed to reduce CO\textsubscript{2} emissions. Many sectors, like the mobility sector, are electrified which leads to an increased electricity demand. There are more loads in the grid whose energy demand needs to be fulfilled to assure a stable operation. Therefore, the power grid, which is already a critical infrastructure, becomes even more important and stability has to be assured at all times.
    
    Realistic power grids have to face a vast amount of possible scenarios, where the system is perturbed from their stable operation. As these perturbations are not necessarily small an investigation beyond linear stability analysis is necessary. Probabilistic methods have proven to be an important tool for studying the stability of power systems. The Basin Stability method, in particular, should be emphasized in this context. It has provided novel insights into multi-stability and resilience for complex systems, but also for power grids in particular. \cite{bs_first, bs_follow_up}. In power grids it is especially interesting to perturb single nodes in the network to find especially vulnerable nodes and to deduce topologies or network motifs which endanger the stability of the network \cite{bs_follow_up, topology_algo}.
    
    Typically, power system stability is divided into several categories, which traditionally include frequency and voltage stability \cite{stability_2021}. The stability of the voltages in power grids is a problem that is likely to become even more relevant in the future \cite{Voltage_Stability} especially due to the increased share of renewables. As renewable energy sources are connected to the grid via inverters, the IEEE has updated their definitions for power grid stability to include converter-driven-stability\cite{stability_2021}, which is mostly concerning inverters in the power grid. This paper will focus on investigating three stability categories, frequency, voltage, and inverter-based stability using probabilistic stability assessments.
    
    Although it is often advocated to increase the share of grid forming inverters, which can maintain grid stability even when no synchronous machines are present, grid following inverters, which introduce constraints into the system and can not sustain grid stability by themselves, are still the majority of inverters today. Therefore, grid following inverters have to be included in a stability assessment. In the case of power grids, the mathematical problem then naturally appears in the form of  differential-algebraic equations (DAEs).
    
    It should be noted that power grids are always subjected to constraints due to Kirchhoff's circuit laws, which ultimately result from fundamental physical laws. The mesh rule is a special case of Faraday's law without an exterior magnetic field and the nodal rule is a result of charge conservation \cite{feynman}. Thus these laws must never be broken and the system reacts infinitely fast to fulfill them. There is also another type of constraint that is less rigid but also important; it arises in the network due to the separation of time scales. 
    
    In case of a perturbation in the power grid the demand can react in a variety of ways, but typically does so much faster than the time scales which we look at and in such a way to recover the energy drawn at the node. Hence we will model the loads as constraints.
    
    Probabilistic stability measures cannot simply be calculated for DAE systems since their phase space is constrained and random valid initial states, i. e. states that fulfill the constraints, cannot be sampled arbitrarily. The process of calculating valid initial conditions for DAEs is called DAE initialization and is a well researched topic in numerical mathematics and considered a challenging numerical problem. 
    Established DAE initialization methods, like Brown's method \cite{DAE_Brown}, can not be used to sample single node perturbations, since only a part of the system is initialized, and it is not possible to regulate what happens to the other variables during the initialization process. Thus it can not be controlled if a perturbation is only targeted to a single node.
    
    In this paper, we will develop a novel method that allows us to sample valid, locally perturbed initial conditions for power grids in DAE form. We will use this method to calculate the Single Node Basin Stability and survivability of an ensemble of power grids and of the IEEE 96 test case. Finally, we will revisit network measures from previous studies and check if they are still a good predictor for the probabilistic stability measures. 

\section{Probabilistic Stability Measures}
    \subsection{Single Node Basin Stability}
    The probability that a power system remains in the operating point and does not run to one of its other attractors depends on its stability against non-small disturbances \cite{bs_first}. The analysis presented by Menck et al. \cite{bs_first} introduced Basin Stability $\langle {\beta}\rangle$ as a non-local and nonlinear measure to quantify the stability of a system.
    The basin of attraction $\mathcal{B}$ of a state $x^*$ is the set of all initial conditions leading back to $x^*$. Analytical computation of basins is challenging and becomes increasingly difficult for complex or higher dimensional systems. Therefore, the authors \cite{bs_first} rely on a single property of the basin $\mathcal{B}$ - its volume. The volume of the basin $\mathcal{B}$ is easier to determine and contains information about the probability of returning to $x^*$. Thus, it can be interpreted as a measure of stability \cite{bs_first}. 
    
    The Basin Stability $\langle {\beta}\rangle$ of a state $x^*$ in a dynamical system is usually calculated by a numerical Monte Carlo method \cite{bs_follow_up}. Disturbed states $x_0$ are randomly drawn from the phase space $\mathcal{S}$. The perturbed system in state $x_0$ is then integrated, and it is checked, whether the system returns to $x^*$ or not.
    The Basin Stability $\langle {\beta}\rangle$ is then defined as:
    \begin{align}
        \langle {\beta}\rangle = \frac{V_{x^*}}{V}
    \end{align}
    where $V$ is the number of all randomly drawn states and $V_{x^*}$ is the number of states that returned to $x^*$.
    In the case of power grids, Basin Stability can be used to study the stability of the system after the disturbance of a single node \cite{bs_follow_up}.
    
    In Single Node Basin Stability, only the variables of a single node are perturbed, while the rest of the nodes remain at the operating point. 
    $\langle {\beta_a}\rangle = [0,1]$ is then the probability that the system returns to the operating point after node $a$ is perturbed.
    
%%%%%%%%%%%%%%%%%%%%%%%%%%%%%%%%%%%%%%%%%%%%%%%%%%%%%%%%%%%%%%%%%%%%%%%%%%%%%%%%%%%%%%%%%%%%%%%%%%    
    
    \subsection{Single Node Survivability}
    Supplementary to the basin stability, which examines the asymptotic behavior of a system after a perturbation, the transient behavior can be analyzed and whether a system will remain in a desirable regime, for some time $t$, after a perturbation. Hellmann et. al. introduces a new stability-related measure, the survivability $\sigma(t)$ of a dynamical system \cite{surv}. The survivability $\sigma(t)$ is the fraction of trajectories that stay within a desirable regime, after an initial perturbations, up to a given time $t$. Again Monte Carlo simulations can be used to calculate $\sigma(t)$, similar to the calculation of the basin stability. Perturbed states $x_0$ are randomly drawn and the perturbed system in state $x_0$ is integrated, and it is checked whether the system leaves the desirable region or not. The survivability is then defined as:
    \begin{equation}
        \sigma(t) = \frac{V_d(t)}{V}
    \end{equation}
    where $V$ again is the number of all randomly drawn states and $V_{d}(t)$ is the number of states which stay inside of the desired region for the simulation time $t$. The single-node survivability is studied by initially perturbing a single node $a$, starting from a stable operating state. $\langle {\sigma_a(t)}\rangle = [0,1]$ is then the probability that the system remains in the desired region, for the time $t$, after node $a$ is perturbed. The desirable frequency region is defined as $\forall  a : |\omega_a | < \omega_d$, the region where the frequency deviations at all nodes stay below a threshold $\omega_d$. The threshold $\omega_{d}$ is chosen identical to the maximal perturbation level of the frequency $\omega_{d} = \Delta \omega$.

    For the voltage condition, we have taken inspiration from the EN 50160 Report \cite{power_quality_standard} for power quality standard\cite{eu_din}. The EN 50160 Report states that the supply voltage variations must remain inside of an interval of $\pm 10\%$ for $95\%$ of a week. We have thus decided to set the bounds for the voltage survivability in such a way that the whole time series is evaluated and it is checked if the voltage of a node remains outside of the $\pm 10\%$ bounds for more than 5 seconds. In this way, we give the system enough time to mitigate the perturbation, which might be bigger than $\pm 10 \%$. 

 %%%%%%%%%%%%%%%%%%%%%%%%%%%%%%%%%%%%%%%%%%%%%%%%%%%%%%%%%%%%%%%%%%%%%%%%%%%%%%%%%%%%%%% 
    
    \subsection{Feasibility}
    Due to the introduction of constraints into the system, there is an additional mechanism that could result in instabilities. In DAEs finite-time solutions exist, even when no singularities are present. The system dynamics can run off to a state where the constraints can no longer be fulfilled. After an initial perturbation to the power grid, it might run into a regime, where the power flow is infeasible, which then results in a system collapse. 
    
    In our Monte-Carlo experiments, we evaluate the constraint equations $g$ at the last time step $t_{end}$ before a system collapse occurs. If the sum of the absolute of all constraints $g_i$ \eqref{eq:infeas} is larger than a threshold $\nu$, we say that the constraints have been violated:
    \begin{equation}
        \sum_i |g_i(z(t_{end}))| > \nu. \label{eq:infeas}
    \end{equation}
    After testing different values, $\nu=0.001$ was selected as the threshold, as it captures the true incidences where $g$ has been violated, but no cases of purely numerical errors in $g$ were counted.
    We introduce the feasibility $f$ as the probability that the power flow will stay feasible after a perturbation:
    \begin{equation}
        f = \frac{V_f}{V}
    \end{equation}
    where $V$ is again the number of all states and $V_f$ is the number of states which stay feasible, according to equation \eqref{eq:infeas}.
     %%%%%%%%%%%%%%%%%%%%%%%%%%%%%%%%%%%%%%%%%%%%%%%%%%%%%%%%%%%%%%%%%%%%%%%%%%%%%%

    \section{Differential-Algebraic Equations}
    \label{sec:dae}
    In this paper we will describe power grids in the form of differential algebraic equations (DAEs). For the next sections we will use the the very general description of explicit DAEs given in equation \eqref{eq:ODE} and \eqref{eq:Constraint}:
    \begin{align}
        \dot{x} &= f(t, x(t),y(t)) \label{eq:ODE} \\ 
        0 &= g(t, x(t), y(t)) \label{eq:Constraint}
    \end{align}
    where equation \eqref{eq:ODE} is an ODE and equation \eqref{eq:Constraint} are the algebraic constraint equations of the system. We will go into details about the modeling choices for power grids in the sections \ref{sec:network_models} and \ref{sec:nodal_models}. 
    
    In the equations \eqref{eq:ODE} and \eqref{eq:Constraint} $x$ depicts the variables whose derivatives $\dot{x}$ appear in the DAE and are called \emph{differential}. The derivatives of the variables $y$ do not appear in the system of equations, i. e. they are referred to as \emph{algebraic}. In the course of the following derivation, an explicit separation between differential and algebraic variables is not necessary, hence we abbreviate $(x,y)$ as $z$. Since all of our equations are autonomous, we can write $g(t, x(t), y(t))$ as $g(z(t))$.
    
    Solving DAE systems is generally much harder than solving pure ODEs because not every point $z$ is a solution of $g$. The problem is hence separated into two parts: first finding valid initial conditions $z_0$, i. e. conditions that fulfill the constraint equations \eqref{eq:Constraint}, and then calculating the trajectories \cite{Geometric_Integration}.
    
    Explicit DAEs can be interpreted as ordinary differential equations, whose solutions lie on a manifold $\mathcal{M}$, which is defined by the constraint equations $g$. Additionally the tangent space at a point $z \in \mathcal{M}$, where $dg(z)$ denotes the total differential is defined as \cite{tangent_space}:
    \begin{align}
        \mathcal{M} &:= \{z \in \mathbb{R}^n: g(z) = 0 \}\\
        T_z\mathcal{M} &= \mathcal{N}(dg(z)). \label{eq:tangent_space}
    \end{align}
    For the scope of this paper, we limit ourselves to smooth manifolds that are embedded in a euclidean space, meaning $\mathcal{M} \subset \mathbb{R}^n$. The Euclidean space which surrounds the manifold $\mathcal{M}$ is called the ambient space $\mathcal{S}$ of $\mathcal{M}$. In the case of power grids, this is not a restriction as they are naturally embedded in an ambient space. Each direction in the ambient space depicts one of the variables $z$, which will become helpful later on.

%%%%%%%%%%%%%%%%%%%%%%%%%%%%%%%%%%%%%%%%%%%%%%%%%%%%%%%%%%%%%%%%%%%%%%%%%%%%%%%%%%%%%%%%% 
    \section{Probability Distributions on Manifolds}
    While the problem of random processes on manifolds is already a long-known issue, for example, Brownian motions on manifolds \cite{brownian_motion}, the authors are not aware that a method to draw random states from a manifold, that are associated with a certain direction of the ambient space, does exist. To calculate 
    probabilistic stability measures of single nodes, these states are needed. Thus we introduce a novel method of sampling states from manifolds, ambient forcing.
    
    \subsection{Ambient Forcing}
    Since the manifolds in our problem are generally unknown, initial conditions can not be directly sampled from $\mathcal{M}$. But the algebraic equations $g(z)$ which define $\mathcal{M}$ and the so-called operation point $z_0$ of the power grid, which lies on $\mathcal{M}$ since it is a fixed point, are known. Starting from the following theorem \ref{theo:SmoothCurves} we will develop our approach.
    \begin{theorem}
    \label{theo:SmoothCurves}
    If $\mathcal{M}$ is a smooth manifold, every vector $v\, \in \, T_z\mathcal{M}$ can be interpreted as the velocity of some smooth curve $\gamma$ in $\mathcal{M}$ \cite{Smooth_Manifold}.
    \end{theorem}
    Therefore if the tangent vectors $v\, \in \, T_z\mathcal{M}$ are known, a relation to the curves in $\mathcal{M}$ exists. Since all manifolds which we are interested in are embedded and smooth, the derivative can be written by using the Jacobi matrix \cite{Smooth_Manifold}. The constraint $g(z(t)) = 0$ is a chained function $g \circ z$ and the chain rule is used to determine the derivative of $g(z(t))$:
    \begin{align}
        \frac{dg(z(t))}{dt} &= J({g \circ z})(t)= J g(z) \cdot \dot z
    \end{align}
    To conserve the constraint equations $g(z(t))$, the derivative must vanish. If $Jg(z(t))$ is known and it is possible to identify its null space $\mathcal{N}(Jg)$, we find the following relation:
    \begin{align}
        \frac{dg(z(t))}{dt} &= J g(z(t)) \cdot \dot z \stackrel{!}{=} \Vec{0}\\
       \Rightarrow \dot z &\in \mathcal{N}(Jg(z(t))). \label{eq:on-surface_velocity}
    \end{align}
    Any trajectory that satisfies equation \eqref{eq:on-surface_velocity} will conserve the constraints $g(z(t_0)) = g(z(t'))$ for all $t' > t_0$ and will thus create states which lie on the manifold if $g(z(t_0))=0$ is satisfied.
    
    Starting from equation \eqref{eq:on-surface_velocity}, we abbreviate the null space of the Jacobian $\mathcal{N}(z) = \mathcal{N}(Jg(z))$ and write $P^\mathcal{N}(z)$ for the matrix representation of the orthogonal projection onto $\mathcal{N}(z)$. Then for an arbitrary dynamic $\dot z = h(z)$, the manifold preserving version can simply be defined as:
    \begin{align}
        \dot z = P^\mathcal{N}(z) h(z) \label{eq: AmbientForcing}
    \end{align}
    Any differential equation that satisfies $h(z) \in \mathcal{N}(Jg(z(t)))$ has trajectories that stay on the manifold if the initial condition $z_0$ lays on the manifold.
    
    In Single Node Basin Stability, a random displacement $F_{rand}$ of the system in the direction of the variables of a node is considered. The directions of the ambient space $\mathcal{S}$ are associated with variables of the nodes. This means that $F_{rand}$ can be drawn from the directions of $\mathcal{S}$ associated with the variables. These variables are now potentially subjected to constraints. A constraint version of such a forcing can still be defined using the definition of equation \eqref{eq: AmbientForcing}:
    \begin{align}
        \dot z = P^\mathcal{N}(z) F_{rand} \label{eq:AmbientForcingFrand}
    \end{align}
    where $F_{rand}$ is an appropriate random, but constant, vector in the ambient space. The definition in equation \eqref{eq:AmbientForcingFrand} is not intrinsic in terms of the constraint manifold $\mathcal{M}$, but depends on the ambient space $\mathcal{S}$ itself.
    
    $F_{rand}$ is a state change corresponding to a failure in the system. The change of state $F_{rand}$ itself is localized at a single node. A distribution of states is obtained, associated with failures localized at single variables or nodes without violating the constraints.
    Both, the amount of $F_{rand}$ and the integration time $t$ of equation \eqref{eq:AmbientForcingFrand} determine the final strength of the random perturbation. A compromise has to be made between sampling strong enough perturbations for a meaningful investigation and to preserve the network locality, which is studied in detail in section \ref{sec:local}. To simplify the dependence of the perturbation strengths we have decided to fix the interval for the random integration times to $[0,1]s$ and to tune the strength of the perturbation by adapting the intervals for the components in $F_{rand}$.
    If a power grid is not constraint in the sense of equation \eqref{eq:Constraint}, meaning that we are in the case of pure ODEs, $g(z)$ is non-existing and we can simply use $z = F_{rand}$, since we do not have to perform a projection.
    %%%%%%%%%%%%%%%%%%%%%%%%%%%%%%%%%%%%%%%%%%%%%%%%%%%%%%%%%%%%%%%%%%%%%%%%%%%%%%%%%%%%%%%%%%%%%%%%%%    
 
    \section{Power Grid Modeling}
    So far we have used the very abstract definition of power grids as DAEs, we want to introduce more specific models which, were used, from the network level \ref{sec:network_models} down to the nodal dynamics \ref{sec:nodal_models}. The nodal dynamics define the equations of the DAEs, while the network gives the connection between them and the power flow. 
    
%%%%%%%%%%%%%%%%%%%%%%%%%%%%%%%%%%%%%%%%%%%%%%%%%%%%%%%%%%%%%%%%%%%%%%%%%%%%%%%%%%%%%%%%%%%%%%%%%%    
    \subsection{Network Models}
    \label{sec:network_models}
    The underlying structure of the power grids will be represented in the form of graphs $\mathcal{G} = \{ N, E \}$, where $\mathcal{G}$ is a set of nodes $N$ and edges $E$ \cite{Newman}. The edges connect the nodes with each other. In power grids, the nodes are the buses and the edges are the transmission lines connecting them. The nodal admittance matrix $Y$ is used in power system analysis to calculate the power flow within the network. It is defined as follows:
    \begin{align}
        Y_{kl} = \begin{cases}
                    y_{k} + \sum_{l = 1, k \neq l}^N y_{kl} & k = l \\
                    -y_{kl} & k\neq l
                \end{cases}
    \end{align}
    where the admittance between two nodes $k$, $l$ is described by  $y_{kl} = g_{kl} + \mathrm{i} b_{kl}$. With $g_{kl}$ being the conductance and $b_{kl}$ the susceptance. The term $y_k$ stands for the admittance of linear loads connected to $k$ and for the admittance-to-ground at $k$ \cite{P_Sauer_book}. The current injected at the nodes is given by Ohms law: $I = Y U$. The nodal Kirchhoff’s law describes the current injections at the nodes: $\sum_j^m i_j = 0$ where $m$ is the total number of all in- and out-coming currents at the node. The Kirchhoff mesh rule is given by: $\sum_j^m u_j = 0$ where $m$ is the number of all voltages around a closed loop. We rewrite the voltages as complex phasors: $u_k = v_k e^{\mathrm{k} \phi_k}$, where $v_k$ is the voltage magnitude and $\phi_k$ is the voltage angle. Using Kirchhoff’s and Ohm’s laws, the power flow of the network is given by equations \eqref{eq:active} and \eqref{eq:reactive} \cite{Power_System_Dynamics}:
    \begin{align}
        S_k &= u_k i_k^* = P_k + \mathrm{i} Q_k \label{eq:apperent}\\
        P_{k} &= \sum_{l=1}^{N} v_k v_j [g_{kl} \cos(\phi_k - \phi_l) + b_{kl} \sin(\phi_k - \phi_l) ]\label{eq:active}\\
        Q_{k} &= \sum_{l=1}^{N} v_k v_l [g_{kl} \sin(\phi_k - \phi_l) + b_{kl} \cos(\phi_k - \phi_l) ]\label{eq:reactive}
    \end{align}
    where $S_k$ is the apparent power at node $k$ and $P_k$ and $Q_k$ are the real and reactive power injected at $k$ respectively.
    
%%%%%%%%%%%%%%%%%%%%%%%%%%%%%%%%%%%%%%%%%%%%%%%%%%%%%%%%%%%%%%%%%%%%%%%%%%%%%%%%%%%%%%%%%%%%%%%%%%%%

    \subsection{Models for nodal dynamics}
    \label{sec:nodal_models}
    \subsubsection{A normal form of grid-forming components}
    \label{sec:Normal_form_voltage}
    In this paper, we will model our dynamical nodes as grid-forming inverters. While synchronous machines and their modeling are a well-researched topic and many textbooks, such as \cite{Power_System_Dynamics, P_Sauer_book}, provide a good overview, such information is still largely missing for grid-forming inverters. While many models to describe grid-forming inverters, such as \cite{dVOC, schiffer_model}, exist unfortunately, a simple, unified model, which still reflects the important properties of grid-forming inverters, does not yet exist.
    
    Recently \cite{normal_form}, R. Kogler et al. have presented a normal form \eqref{eq:normalform} that can capture the most basic nonlinearities of various power system dynamics but especially for grid forming components. The normal form is given as:
    \begin{align}
        v_k^2 &= u_k u_k^* \nonumber \\
        \frac{d\omega_k}{dt} &= (A_{x,k} + B_{x,k} \omega_k + C_{x,k}  v_{k}^2 + G_{x,k} P_k + H_{x,k}  Q_k) ) \frac{1}{M_{x,k}} \label{eq:normalform} \\
        \frac{du_k}{dt} &= (A_{u,k} + B_{u,k} \omega_k + C_{u,k} v_{k}^2 + G_{u,k}  P_k + H_{u,k} Q_k ) \cdot u_k \nonumber
    \end{align}
    where $\omega_k$ and $u_k$ are the frequency and voltage at node $k$ respectively, $P_k$ and $Q_k$ represent, as in equation \eqref{eq:apperent}, the active and reactive power at $k$ and $v_k$ is the voltage magnitude. All the other coefficients are modeling parameters.
    
    In \cite{normal_form} the authors have demonstrated that a large number of established models, such as in \cite{dVOC} and \cite{schiffer_model}, can be mapped to the normal form. Furthermore, the normal form was validated by numerical simulations as well as by experimental data. Therefore it is the perfect candidate for the nodal dynamics of the grid-forming inverters in this work. As the derivatives of the voltage and frequency show up in the normal form model we will also refer to them as \emph{differential} nodes, following the definition in section \ref{sec:dae}.
    
 %%%%%%%%%%%%%%%%%%%%%%%%%%%%%%%%%%%%%%%%%%%%%%%%%%%%%%%%%%%%%%%%%%%%%%%%%%%%%%%%%%%%   
    \subsubsection{PQ Loads}
    \label{sec:loads_constraints}
    The consumers in this paper will be modeled by the traditional PQ load model. We could have chosen other models, such as the exponential recovery load, which reacts to the applied voltage but is nevertheless a constraint. However, the focus of this work is on the study of the effect of constraints on power systems, so it is reasonable to choose the simplest static model. The PQ node locally fixes the power output of node $k$:
    \begin{align}
        0 = (P_k + \mathrm{i} Q_k) - u_k \cdot i^{*}_k \label{eq:PQNode}
    \end{align}
    The voltage $u_k$ and current $i_k$ at node $k$ are now subjected to the constraint equation. As the derivatives of the voltage do not show up in the PQ model, we will also refer to them as \emph{algebraic} nodes.
    
    %%%%%%%%%%%%%%%%%%%%%%%%%%%%%%%%%%%%%%%%%%%%%%%%%%%%%%%%%%%%%%%%%%%%%%%%%%%%%%%%%%%%%%%%%%%%%%%%%%%%%%55
    
    \subsection{Network Ensemble}
    To apply our method and explore the probabilistic stability of power grids, an ensemble of 100 synthetic power grids is drawn by using a random growth algorithm \cite{pg_growth_model}. This algorithm generates synthetic networks which share their topological properties with realistic power grids when a set of appropriate parameters is chosen.
    
    All of the networks have 100 nodes each, which are randomly turned into PQ nodes \eqref{eq:PQNode} or into normal forms \eqref{eq:normalform}. Additionally, the nodes are randomly turned into consumers or producers by drawing a random power output between $[-1,1] p.u.$ for each node.
    
    PQ nodes with a positive power output depict grid feeding inverters. When their power output is negative, they represent traditional loads. Normal forms with positive power output depict grid forming inverters. Those with a negative power output depict aggregated loads, for example, a small village, which produces solar power and is connected to the next higher grid layer via a grid forming inverter but ultimately consumes more power than it can produce.
    
    In the investigation in section \ref{sec:local} random frequency deviation in $[-5, 5]$ and voltage perturbations in $[-2, 2]$ are sampled using Ambient Forcing. Starting from section \ref{sec:topological} the frequencies of the nodes were randomly perturbed in $[-3, 2]$ and the voltages in $[-2, 0]$, which depicts a probabilistic version of a short circuit. We used $V=250$ initial states for each node in the network.
    
%%%%%%%%%%%%%%%%%%%%%%%%%%%%%%%%%%%%%%%%%%%%%%%%%%%%%%%%%%%%%%%%%%%%%%%%%%%%%%%%%%%%%%%%%%%%%%%%%%%%%%%%%%%%%
    
\section{Localized Perturbations}
    \label{sec:local}
    Ambient Forcing is intended to be network local, meaning that the initial perturbation only targets a single node, but is not clear if it succeeds. Unlike in ODEs, in DAE systems localized perturbations are not mathematically well defined, since not all initial conditions are valid. We have to project the localized perturbation vectors $F_{rand}$ from the ambient space $\mathcal{S}$ onto the manifold.
    
    In the case of power grids, this reflects a real-life behavior of the constraints. When a constraint voltage $u_a$ is perturbed it might not be possible for the node $a$ itself to fulfill the constraint as the current $i_a$ can not be altered freely since it is additionally constraint as it has to fulfill Kirchhoff's. Therefore changing the voltage $u_a$ results in a voltage shift at adjacent nodes, which results in an alerted current $i_a$, to fulfill the desired power output.

    In power grids, the perturbations will naturally spread to adjacent nodes when constraints are present. We will characterize the network locality of the random distribution of perturbations $\rho_k$ at node $k$ by introducing a novel network measure that is related to the closeness centrality.
    
    First the amplification $A(k)$ of an node $k$ is defined as the sum norm of the mean displacement vector from the operation point $z_0$ $\Delta z(k) = z_0 - z_{rand}(k)$ after a perturbation on node $k$. The amplification $A(k)$ is thus the mean total strength of a perturbation. $A(k)$ is used to calculate the perturbation fractions $p_a(k)$ of all nodes $a$ in the network:
    \begin{align}
        A(k) &= \sum_l |\Delta z_l(k)| \\
        p_l(k) &= \frac{|\Delta z_l(k)|}{A(k)}
    \end{align}
    The \textit{spreadability} $s(k)$ is then defined as the sum over all fractions $p_l(k)$ weighed by the shortest path length $d(k,l)$ between the perturbed node $k$ and all the other nodes $l$. A slightly modified form of the normalized closeness centrality $C_c(k)$ \cite{closeness_centrality} is used to normalize $s(k)$. The number of all nodes $n$ is used as the numerator in the modified version $C'_c(k)$ instead of the usual $n - 1$:
    \begin{align}
        C'_c(k) &= \frac{n}{\sum_a d(k,l)}\\
        s(k) &= C'_c(k) \sum_l p_l(k) d(k,l). \label{eq:spreadability}
    \end{align}
    If only node $k$ is displaced the spreadability vanishes. If all nodes in the network are equally perturbed then $s(k) = 1$. The spreadability of each node in the ensemble was calculated using equation \eqref{eq:spreadability}. The data of the network ensemble is categorized by the minimal constraint distance $\min(d_c)$. Constraint nodes themselves have a minimal constraint distance of 0. The data is also sorted by the minimal distance to the \emph{next constraint} $\min(d_{cn})$, the type of the node itself is not considered. This means that neighbors of constraint nodes will always have a distance $\min(d_{cn})$ of 1, independently of their type.
    
\begin{figure}%[H]
            \begin{subfigure}[b]{0.49\textwidth}
            \includegraphics[width=\textwidth]{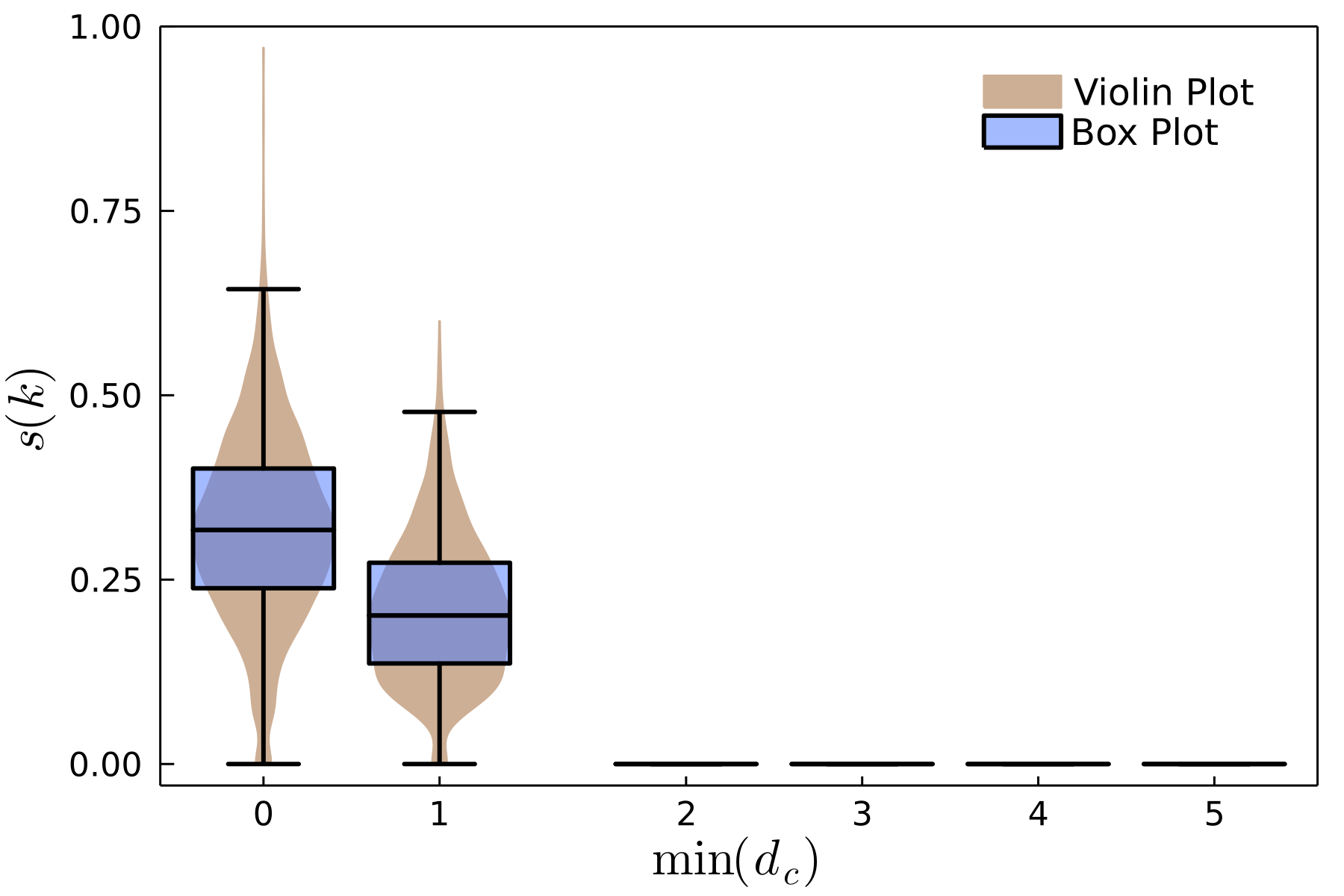}
            \caption{}
            \label{fig:spread_dc}
        \end{subfigure}
        \hfill
        \begin{subfigure}[b]{0.49\textwidth}
            \includegraphics[width=\textwidth]{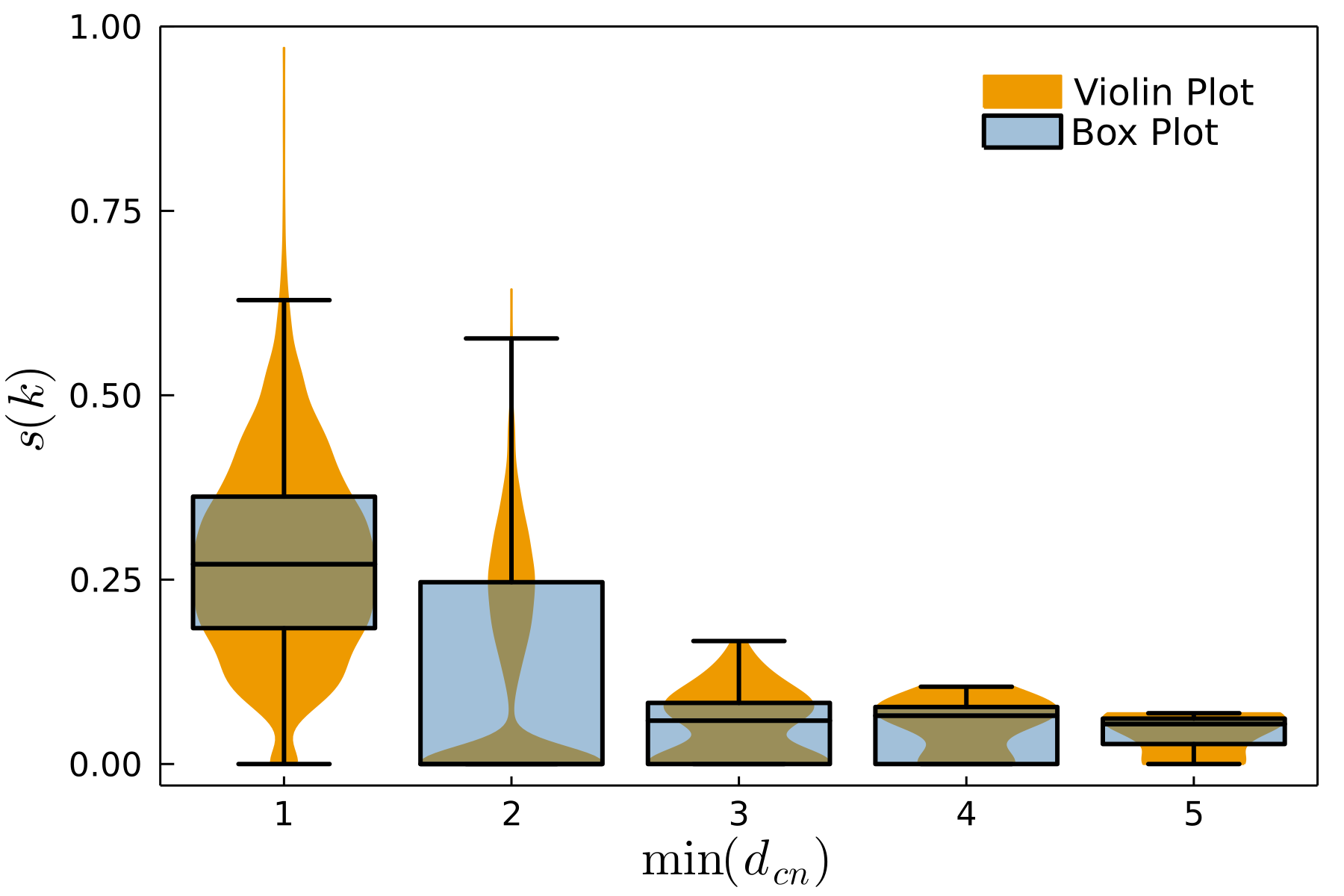}
            \caption{}
            \label{fig:spread_dcn}
            \end{subfigure}
            \caption{Box plots of the spreadabilities using different sorting mechanisms for the distance. Figure \ref{fig:spread_dc} shows that only constraint nodes or neighbors of constraint nodes spread. The second plot \ref{fig:spread_dcn} shows that the spreadability of the nodes decreases with the distance to the next constraint in the system, independently if the node itself is a constraint or not.}
        \end{figure}

    The data is represented by a combination of a box and violin plot shown in the figures \ref{fig:spread_dc} and \ref{fig:spread_dcn}. If the ambient forcing algorithm does not introduce any further spreading of perturbations only constraints and nearest neighbors of constraints will have a non-vanishing spreadability. This behavior can be observed in figure \ref{fig:spread_dc}, meaning that the ambient forcing algorithm can be used for single node perturbations. Furthermore, in figure \ref{fig:spread_dcn}, it can be seen that the spreadability decreases as the distance to the next constraint in the power grid increases, independently of the node type.
    
%%%%%%%%%%%%%%%%%%%%%%%%%%%%%%%%%%%%%%%%%%%%%%%%%%%%%%%%%%%%%%%%%%%%%%%%%%%%%%
\section{Results}
\subsection{Topological Properties}
    \label{sec:topological}
    The algorithm, which was introduced in \cite{topology_algo}, classifies nodes according to their topological properties and can identify vulnerable nodes in power grids. In this algorithm nodes with degree $d=1$ are iteratively removed, introducing a sub-graph at each level. 
    
    We used the topological classiﬁcation scheme to investigate whether it is still possible to separate the nodes according to their stability properties when constraints, as well as the normal form, are introduced into the network. While we can report that the classes are still able to separate the nodes with respect to their stability properties, we have found that the degree is a much better predictor than the topological classes \cite{topology_algo}, for both $\sigma$ and $\beta$ but especially for the survivability.
    
\begin{figure}%[H]
        \centering
            \begin{subfigure}[b]{0.49\textwidth}
                \includegraphics[width=\textwidth]{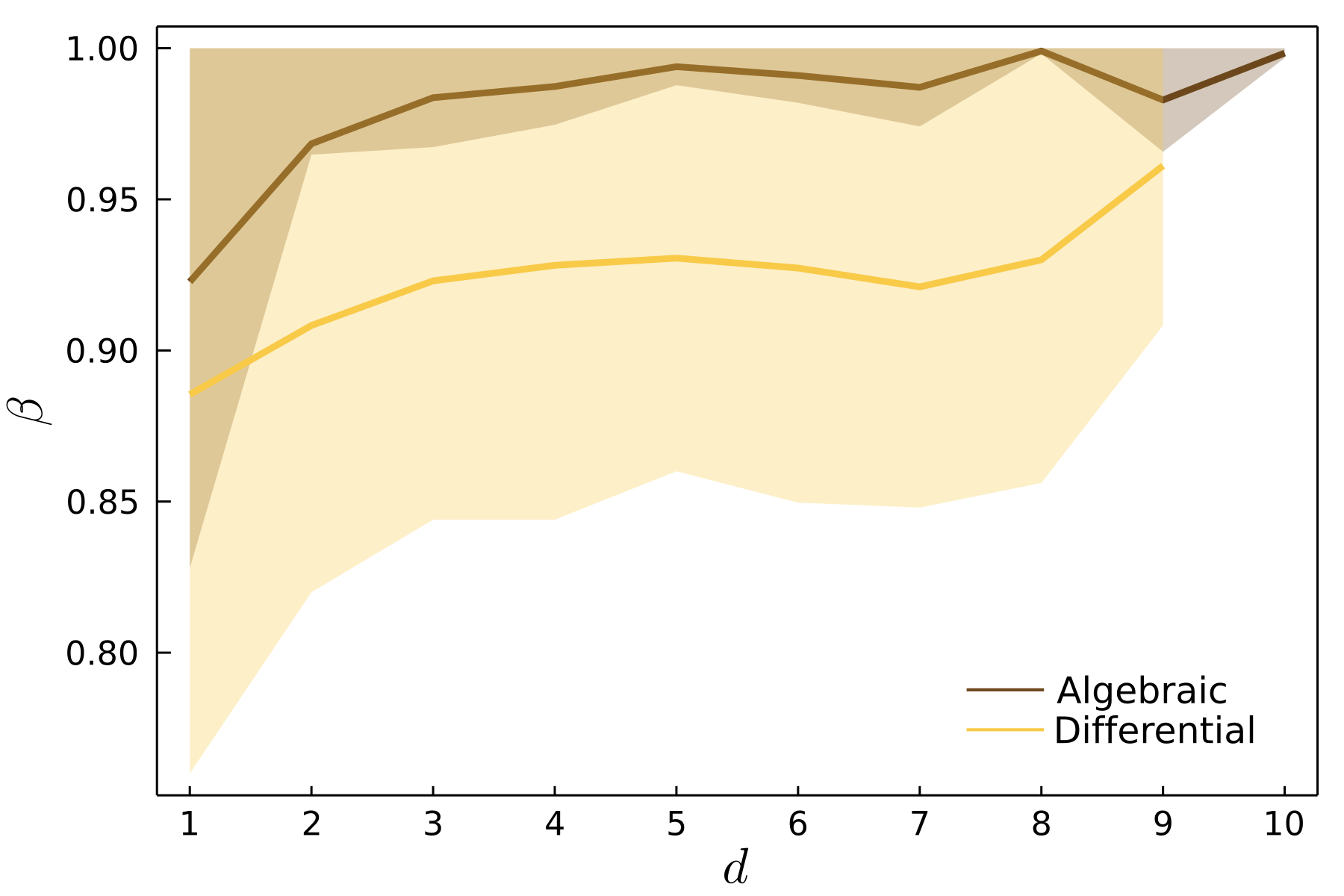}
                \caption{Basin Stability}
                \label{fig:degree_snbs}
            \end{subfigure}
            \hfill
            \begin{subfigure}[b]{0.49\textwidth}
                \includegraphics[width=\textwidth]{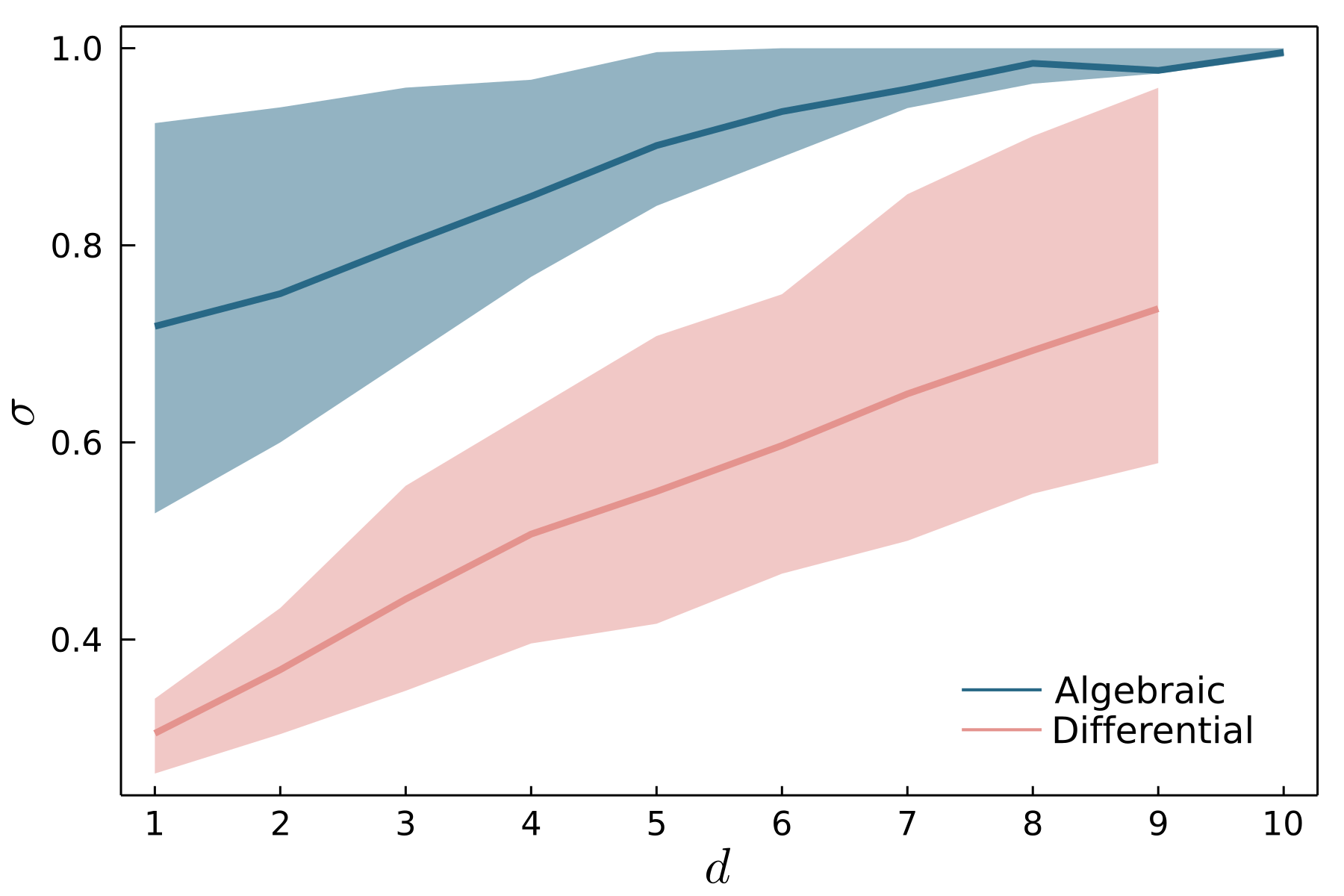}
                \caption{Survivability}
                 \label{fig:degree_surv}
            \end{subfigure}
        \caption{Degree dependence of the basin stability \ref{fig:degree_snbs} and survivability \ref{fig:degree_surv}. Solid lines represent the mean and the shaded areas depict the $15.9 \%$ and $84.1 \%$ percentile. Both the basin stability and survivability increase with the degree.}
    \end{figure}

    To illustrate the connection between the degree and the probabilistic measures, we have plotted the mean $\beta$ and $\sigma$ with respect to the degree in the figures \ref{fig:degree_snbs} and \ref{fig:degree_surv}. The solid lines represent the mean and the shaded area depicts the $15.9 \%$ and $84.1 \%$ percentile, which depict one standard deviation in the normal distribution.
    
    The basin stability, in figure \ref{fig:degree_snbs}, shows a weak upwards trend regarding the degree. 
    The picture for the survivability on the other hand clearly exhibits that for both the algebraic and differential nodes the mean survivability increases with the degree. This is in stark contrast to the finding of the original paper \cite{topology_algo}, where the survivability decreases with the degree. For the algebraic nodes, we see that the size of the percentiles decreases with the degree, while it increases with the degree for the differential nodes. 
    
    We have verified that the strength of the mean perturbations $A$ does not decrease with the degree. The results of this investigation can be found in the appendix \ref{sec:amp}.
    
    % Degree

\begin{figure}%[H]
        \centering
            \begin{subfigure}[b]{0.49\textwidth}
                \includegraphics[width=\textwidth]{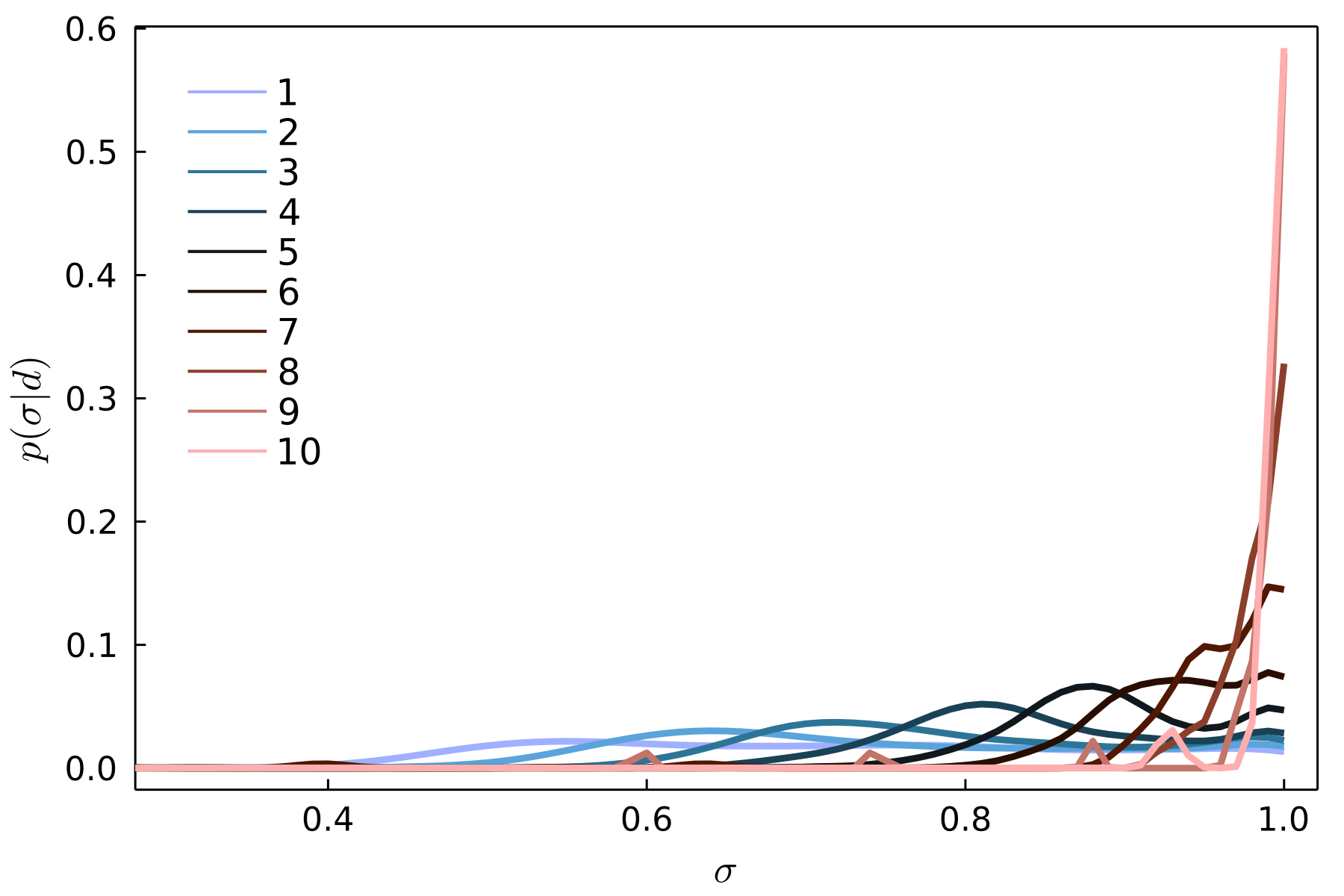}
                \caption{Algebraic}
                \label{fig:dens_surv_cons}
            \end{subfigure}
            \hfill
            \begin{subfigure}[b]{0.49\textwidth}
                \includegraphics[width=\textwidth]{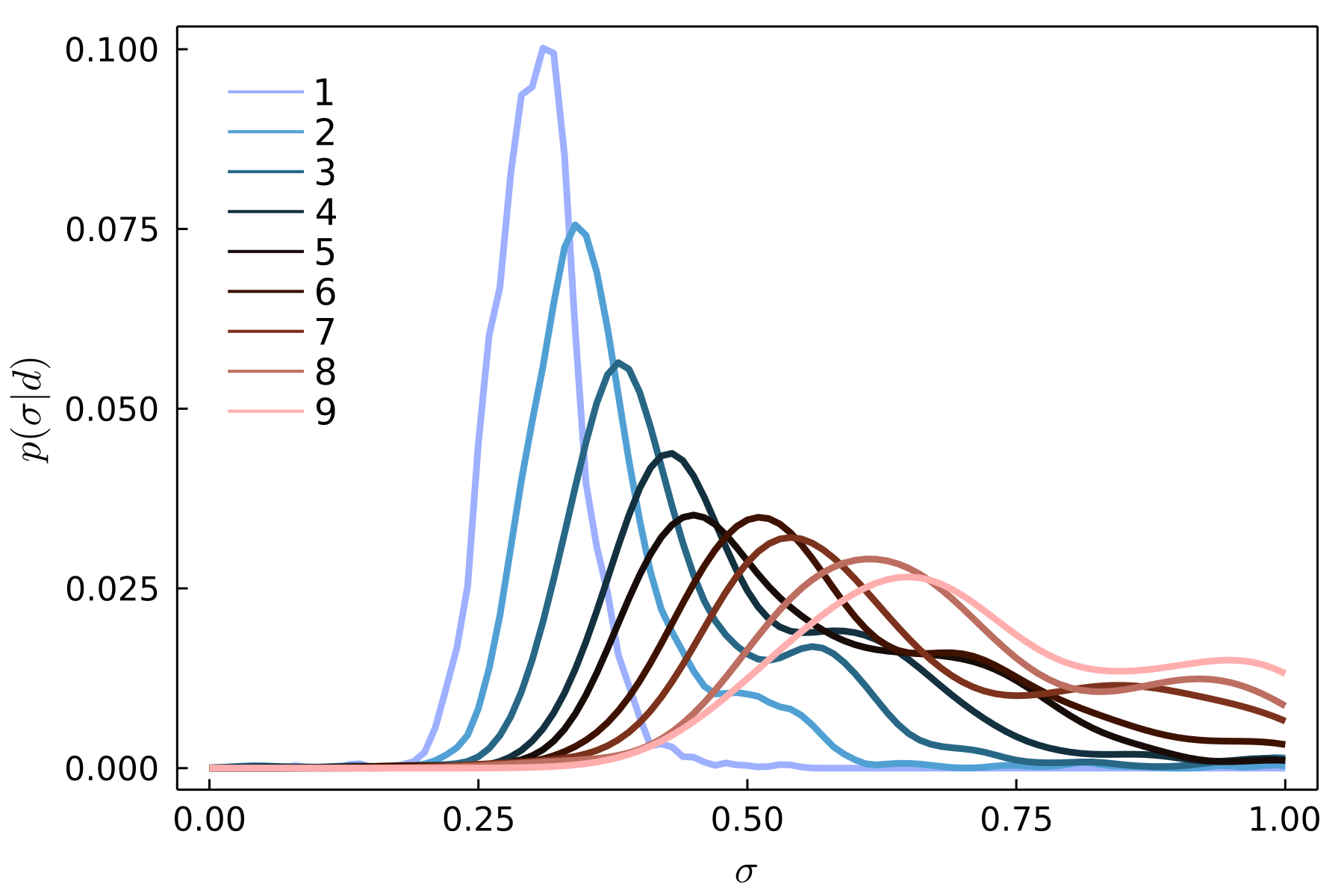}
                \caption{Differential}
                 \label{fig:dens_surv_dyn}
            \end{subfigure}
        \caption{Density Plot of the survivability classified by the degree of the nodes. The left figure shows all algebraic nodes, while the differential nodes are shown on the right. It can be seen that the survivability increases with the degree, especially for the differential nodes.}
    \end{figure}

    As the effect of the degree on the survivability is especially prominent, we have investigated this behavior further by calculating the probability densities of the survivability for all nodes in the ensemble.
    For the algebraic nodes, whose results are given in Fig. \ref{fig:dens_surv_cons}, we find that nodes with a degree of 8 or higher survive most often and their mode is $\sigma = 1$. Nodes with smaller degrees survive less frequently and the distributions start to broaden. 
   
    The survivability of the differential nodes, shown in figure \ref{fig:dens_surv_dyn}, exhibits the most striking dependence on the degree. Increasing the degree in the nodes results in a higher mode of the survivability. The distribution of $\sigma$ becomes broader and more long-tail with an increasing degree. This is in contrast to the behavior of the algebraic nodes where the distributions become sharper with an increasing degree.
    
    Additionally, we want to draw attention to the fine structuring, which becomes visible in the survivability of the differential nodes. Already starting at a degree of $d=2$, a second peak with higher survivability emerges. This hints that there might be a second effect that influences the survivability.
    
%%%%%%%%%%%%%%%%%%%%%%%%%%%%%%%%%%%%%%%%%%%%%%%%%%%%%%%%%%%%%%%%%%%%%%%%%%%%%%%%%%%

    \subsection{Types of Instabilities in the System}
    Different possible mechanisms could result in instabilities and it is worthwhile investigating which of these mechanisms are most important. The following classification for instabilities will be used. First, we count events where the frequency of at least one node does not return to the synchronous state, i. e. that the system has desynchronized. Additionally we compute the states where the voltage of at least one node does not return to the fixed point. We will also fine-grain these and measure the events where \emph{only} the frequency is desynchronized and then measure the states where only the voltage runs off to another fixed point. Finally, we will look at all the nodes where infeasible power flows have occurred.
    
\begin{figure}[H]
            \centering
            \begin{subfigure}[b]{0.49\textwidth}
                \includegraphics[width=\textwidth]{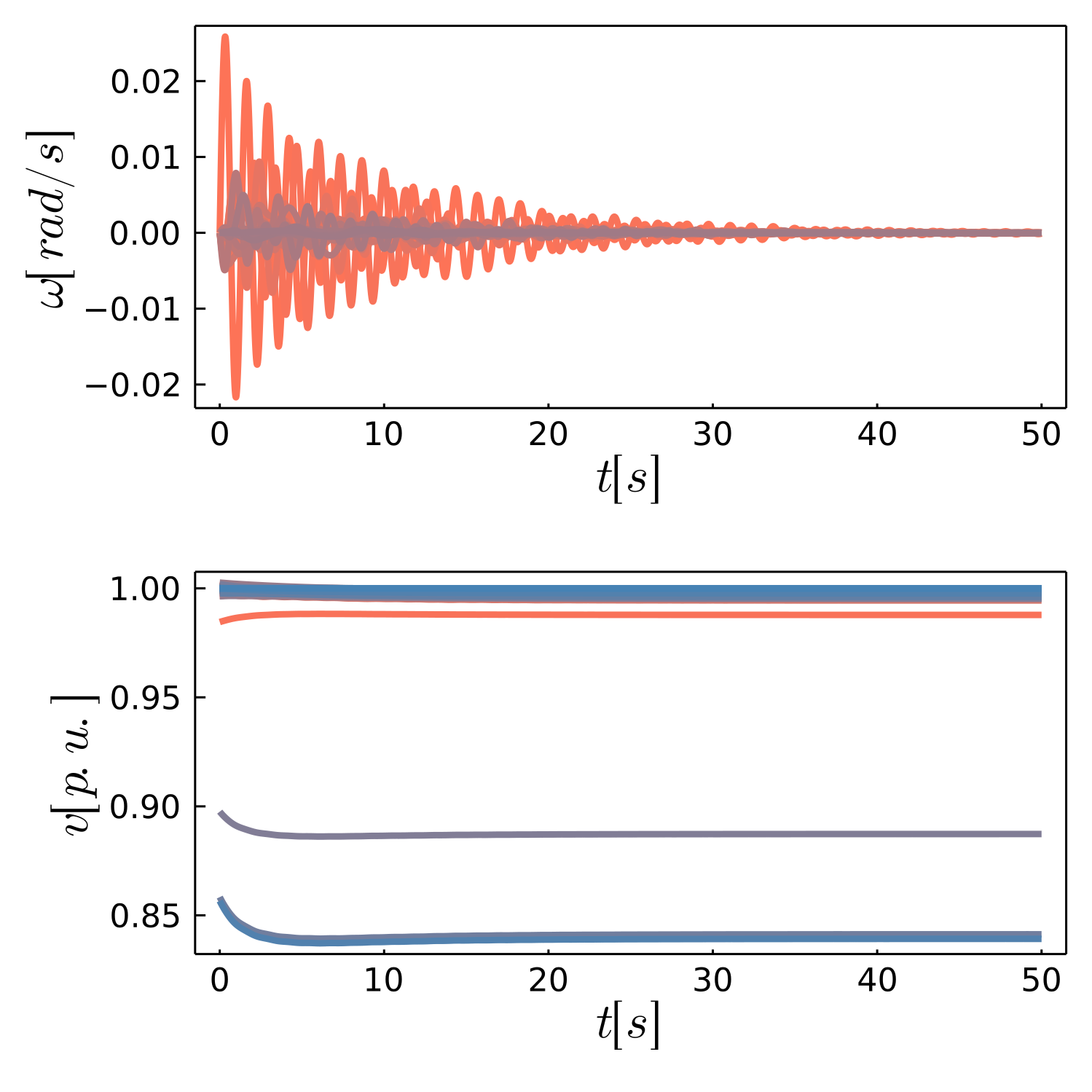}
                \caption{Node 94}
            \end{subfigure}
            \hfill
            \begin{subfigure}[b]{0.49\textwidth}
                \includegraphics[width=\textwidth]{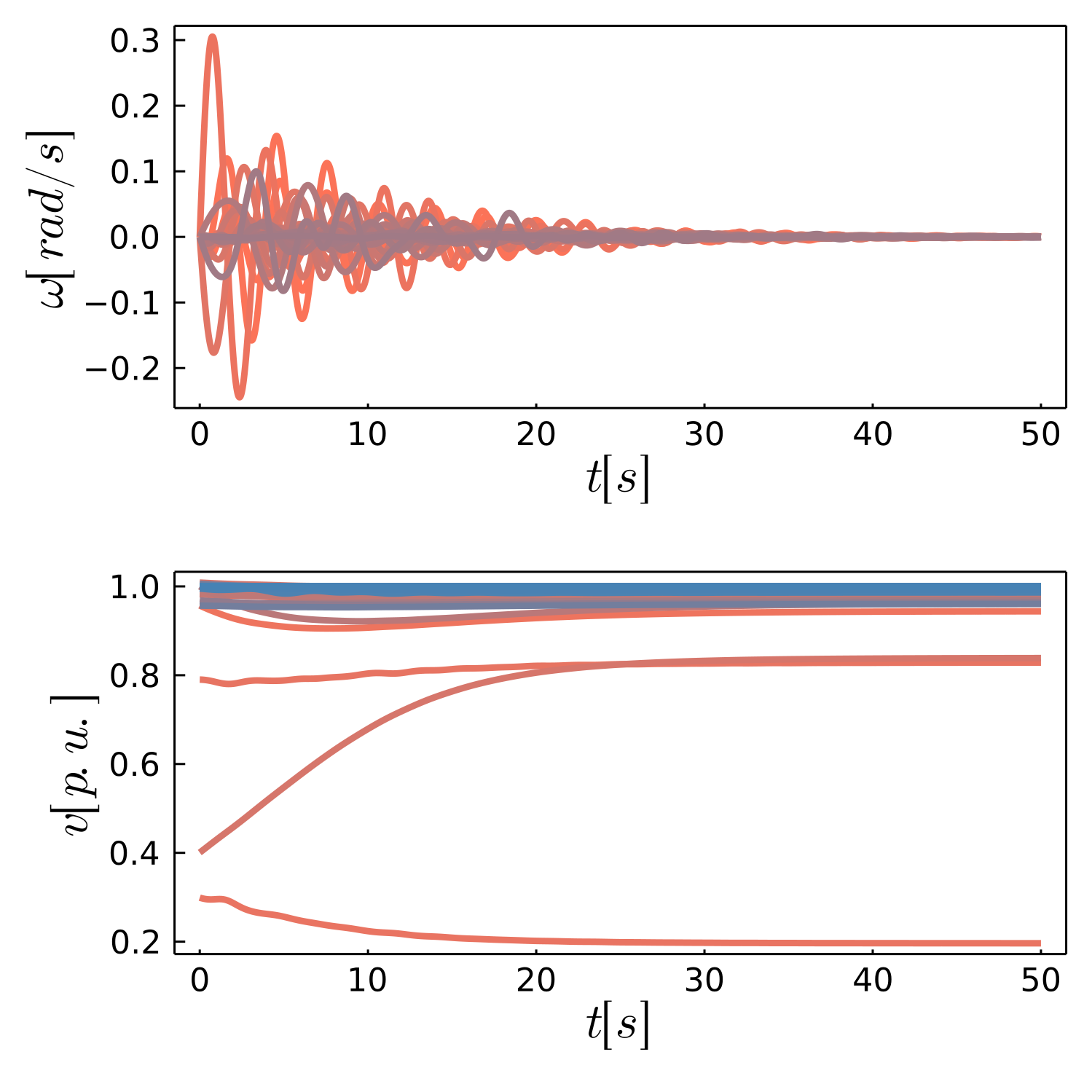}
                \caption{Node 13}
            
            \end{subfigure}
        \begin{subfigure}[b]{0.49\textwidth}
                \includegraphics[width=\textwidth]{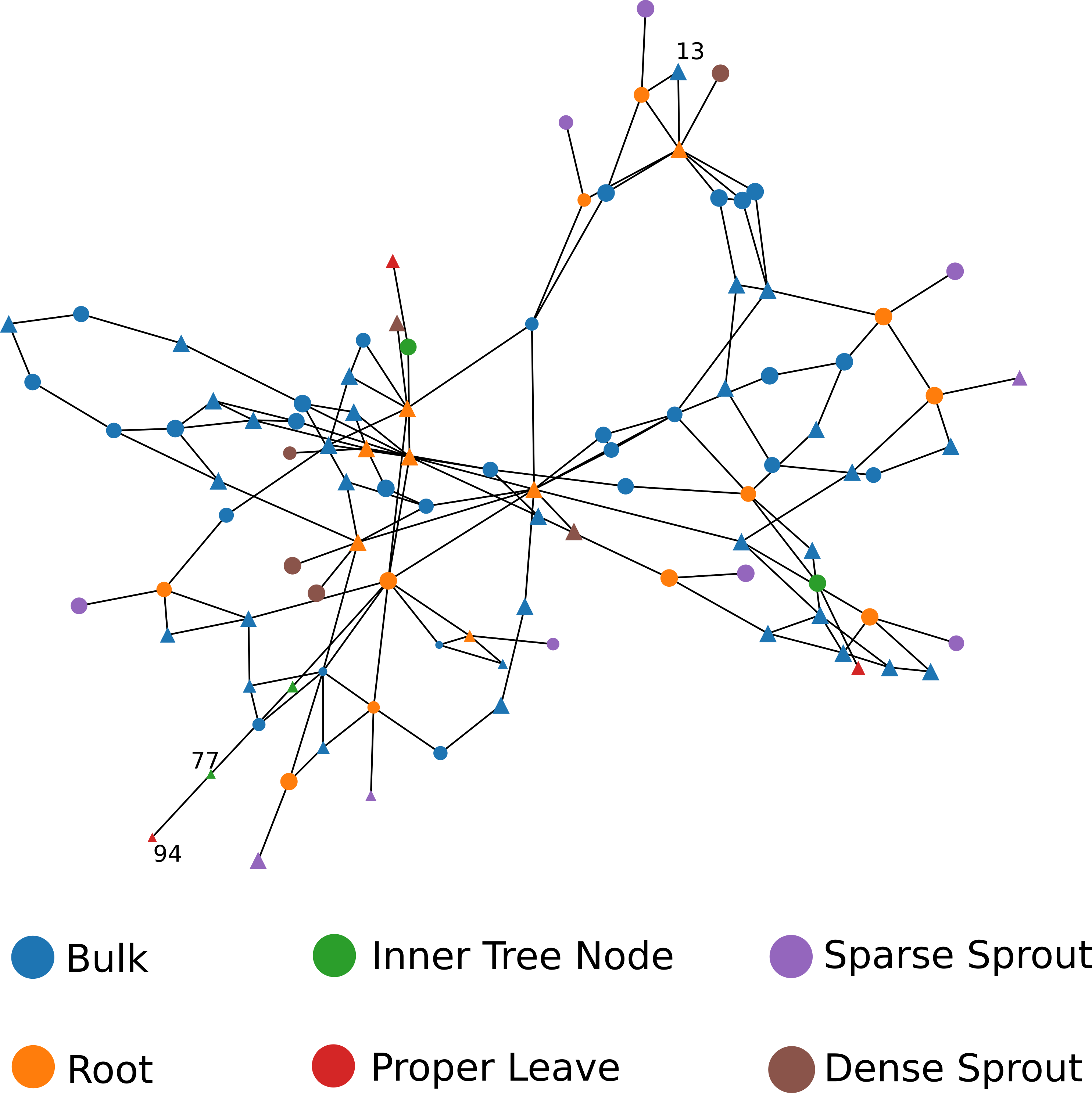}
        \caption{Power Grid}
        \end{subfigure}
        \caption{Example of perturbations on nodes that lead to voltage drops without a frequency desynchronization. Node 77, which is also labeled, runs to the same fixed point as node 94 after a perturbation.}
        \label{fig:vol_drop_only}
    \end{figure}

    We investigate the percentage of nodes $p_n$ in the ensemble, where these events have occurred at least once. The results are shown in the table \ref{tab:node_fault_events}. We can see that the desynchronization only occurs at 19.1$\%$ of the nodes at all, and never occur on their own, without a change in voltage. Two example trajectories of these desynchronized states can be found in the appendix \ref{fig:desync_vol_drop}. Voltage drops occur at 29.5$\%$ of the nodes in the network and also appear without a desynchronization, at 11.4$\%$ percent of the nodes. Two examples of voltage drops that occur without desynchronization can be found in the figures \ref{fig:vol_drop_only}. Finally, infeasible power flows appear most commonly, at 37.1$\%$ of the nodes.
    
    \begin{table}[H]
        \centering
        \begin{tabular}{|l|l|l|l|l|}
        \hline Faults & Desynchronization  & Voltage Drops & Only Voltage Drops  & Infeasible Flows\\ 
        \hline $p_n$ & 19.1 $\%$ & 29.5 $\%$ & 11.4 $\%$ & 37.1 $\%$    \\ \hline
        \end{tabular}
        \caption{Percentage of nodes $p_n$ in the ensemble where faults have appeared at least once.}
        \label{tab:node_fault_events}
    \end{table}
    As the infeasible power flows occur most commonly we have decided to compare the feasibility $f$ to the basin stability and survivability. For this purpose, we have calculated the probability densities of $\beta$, $\sigma$, and $f$ for the entire ensemble and compared the distributions using the Jensen-Shannon divergence\cite{jsd}, which is a symmetric and bounded version of the Kullback-Leibler divergence $D_{KL}(P,Q)$ \cite{kullback_leibler}. $D_{KL}(P,Q)$ is a statistical distance measure between two probability distributions $P$ and $Q$, which are defined in the same probability space $X$:
    \begin{align}
        D_{KL}(P,Q) = \sum_{x\in X} P(x) \log \left( \frac{P(x)}{Q(x)} \right)
    \end{align}
    and it is a measure for the information which is lost if we approximate $P(X)$ using $Q(X)$.
    The Jensen-Shannon-divergence for $P(X)$ and $Q(X)$ is then defined as:
    \begin{align}
        JSD(P, Q) &= \frac{1}{2} D_{KL}(P,M) + \frac{1}{2} D_{KL}(Q,M)\\
        M &= \frac{1}{2} (P(X) + Q(X)).
    \end{align}
    The Jensen-Shannon-divergence is bounded on the interval $[0,1]$ and $JSD(P,Q)$ is 0 if $P(X)$ and $Q(X)$ are equal and 1 if both probability distributions are not correlated.
    
    The Jensen-Shannon-divergence can be calculated for $\beta$, $\sigma$, and $f$ as they are all defined in the same probability space and the possible outcomes $x\in X$ are defined on the interval of $[0,1]$. The Jensen-Shannon divergence $JSD(\beta, f)$ for the algebraic nodes is 0.079 and for the differential nodes, it is 0.013. $JSD(\sigma, f)$ is 0.15 for the algebraic nodes and 0.248 for the differential nodes. This means that little information is lost if we use $f$ to approximate $\beta$, while a lot of information is lost if $f$ is used to approximate $\sigma$. From this, we can conclude that the infeasible power flows are the driving mechanism that lead to basin-unstable systems, while other effects dominate the survivability.
    
    Finally, we can also infer that there are no differences between producers or consumers, i.e. the sign of the power, concerning the basin stability or survivability. This means that not only inverters are at risk of inducing voltage drops or infeasible power flows but also loads. Therefore it is necessary to pay close attention to all types of actors in the grid. 
    
    From this, we can conclude that although algebraic nodes themselves are more stable and survive more frequently than differential nodes, which can be seen in the figures \ref{fig:degree_snbs} and \ref{fig:degree_surv}, they are responsible for most of the instabilities in the grid, the infeasible power flows. This is another argument for pushing ahead with the expansion of grid-forming inverters in order not to increase the number of constraints in the grid even further, since loads, which introduce the same problems, will always be connected to power grids.

    A more detailed load model is needed to properly understand this phenomena but a fast energy storage at the loads could be used to change the power flows in time and mitigate the infeasabilities.
    In a future investigation \emph{Monte Carlo basin bifurcation analysis} \cite{mcbb} could be used to investigate the different asymptotic states in such systems and their basin of attraction.
    
    %%%%%%%%%%%%%%%%%%%%%%%%%%%%%%%%%%%%%%%%%%%%%%%%%%%
    \section{IEEE 96 System}
    Finally, we will also apply our new method to a more realistic test case. The IEEE 96 test system \cite{ieee96} was designed for bulk power system reliability evaluation studies. This test system is not representative of any specific or typical power system, as it should hold universal characteristics to be useful as a reference for testing the impact of different evaluation techniques. The system should thus be seen as a hybrid and atypical system. To update the system and show the influence of inverters, we have decided to model the generators as normal forms \eqref{eq:normalform}. The loads are still depicted by traditional PQ nodes \eqref{eq:PQNode}. The parameters used for the normal form can be found in the appendix \ref{sec:parameter}. 
    
 %%%%%%%%%%%%%%%%%%%%%%%%%%%%%%%%%%%%%%%%%%%%%%%%%%%%%%%%%%%%%%%%%%%%%%%%
    \subsection{EN 50160 Report Perturbations}
    The perturbation parameters were chosen following the EN 50160 Report \cite{power_quality_standard} on Power Quality Standards. In this standard, the grid frequency $\omega$ must remain inside of the bounds $\Delta \omega = [-3.0, 2.0]$ for 100$\%$ of the observed time. The random frequency perturbations are thus sampled from this interval and the thresholds for the survivability are chosen identical. We have decided to first perturb the voltage of each node in the network between $\pm 10\%$ from its operation point level. For the voltage survivability, it is checked whether the voltage of a node remains outside of $\pm 10\%$ of the operation point level for more than 5 seconds. 
    
    These results show that the minimal basin stability of a node is 0.983$\pm$ 0.006 and the minimal survivability is 0.95 $\pm$ 0.01. Thus it was decided to not include a more extensive evaluation of the data as there is little additional information that can be gathered, since all nodes are almost globally stable. Nevertheless, it is an interesting result as it demonstrates that the IEEE test case with inverters fulfills the requirements of the EN report. 
    
%%%%%%%%%%%%%%%%%%%%%%%%%%%%%%%%%%%%%%%%%%%%%%%%%%%%%%%%%%%%%%%%%%%%%%%%
    \subsection{Increased Perturbation Strength}
\begin{figure}%[H]
        \centering
            \begin{subfigure}[b]{0.49\textwidth}
                \includegraphics[width=\textwidth]{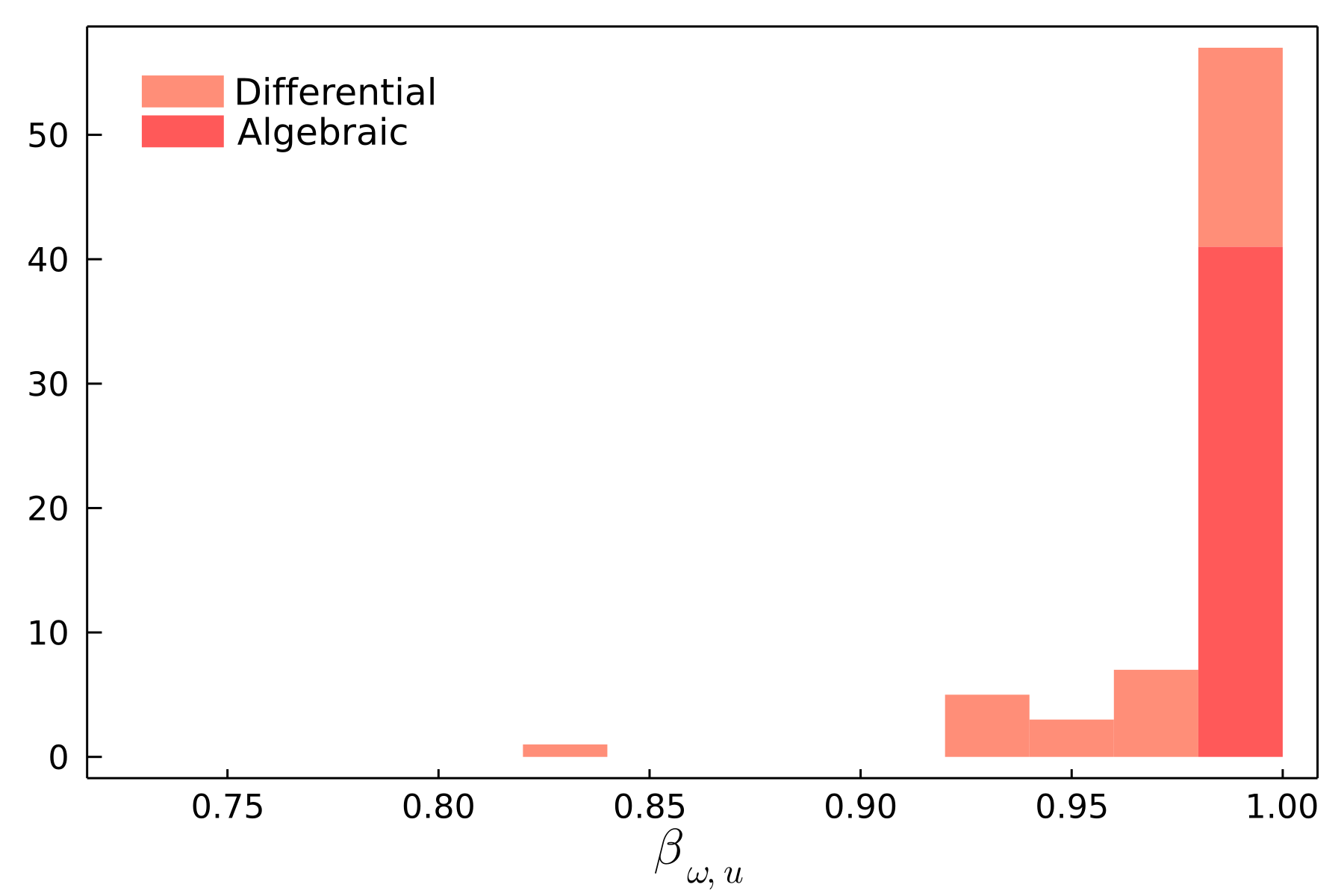}
                \caption{Basin Stability}
                \label{fig:hist_96Bus_snbs_5}
            \end{subfigure}
            \hfill
            \begin{subfigure}[b]{0.49\textwidth}
                \includegraphics[width=\textwidth]{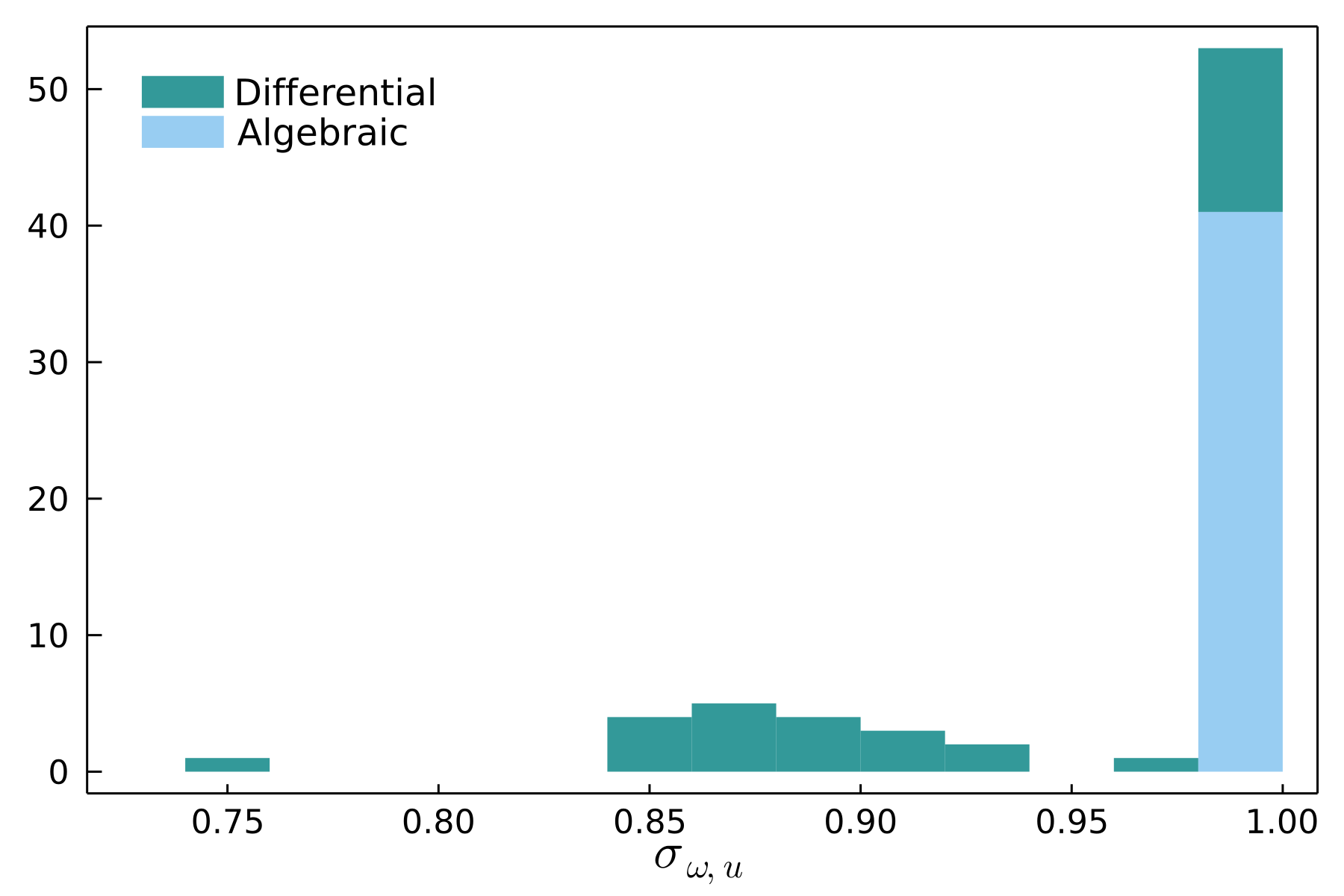}
                \caption{Survivability}
                 \label{fig:hist_96Bus_surv_5}
            \end{subfigure}
        \caption{Stacked Histograms of the Basin Stability \ref{fig:hist_96Bus_snbs_5} and survivability \ref{fig:hist_96Bus_surv_5}, grouped by the function type. The basin stability of differential and algebraic nodes are similarly distributed. The second figure shows a clear distinction in the distribution of the survivability between differential and algebraic nodes.}
    \end{figure}

\begin{figure}[H]
       \centering
       \includegraphics[width = 0.5\textwidth]{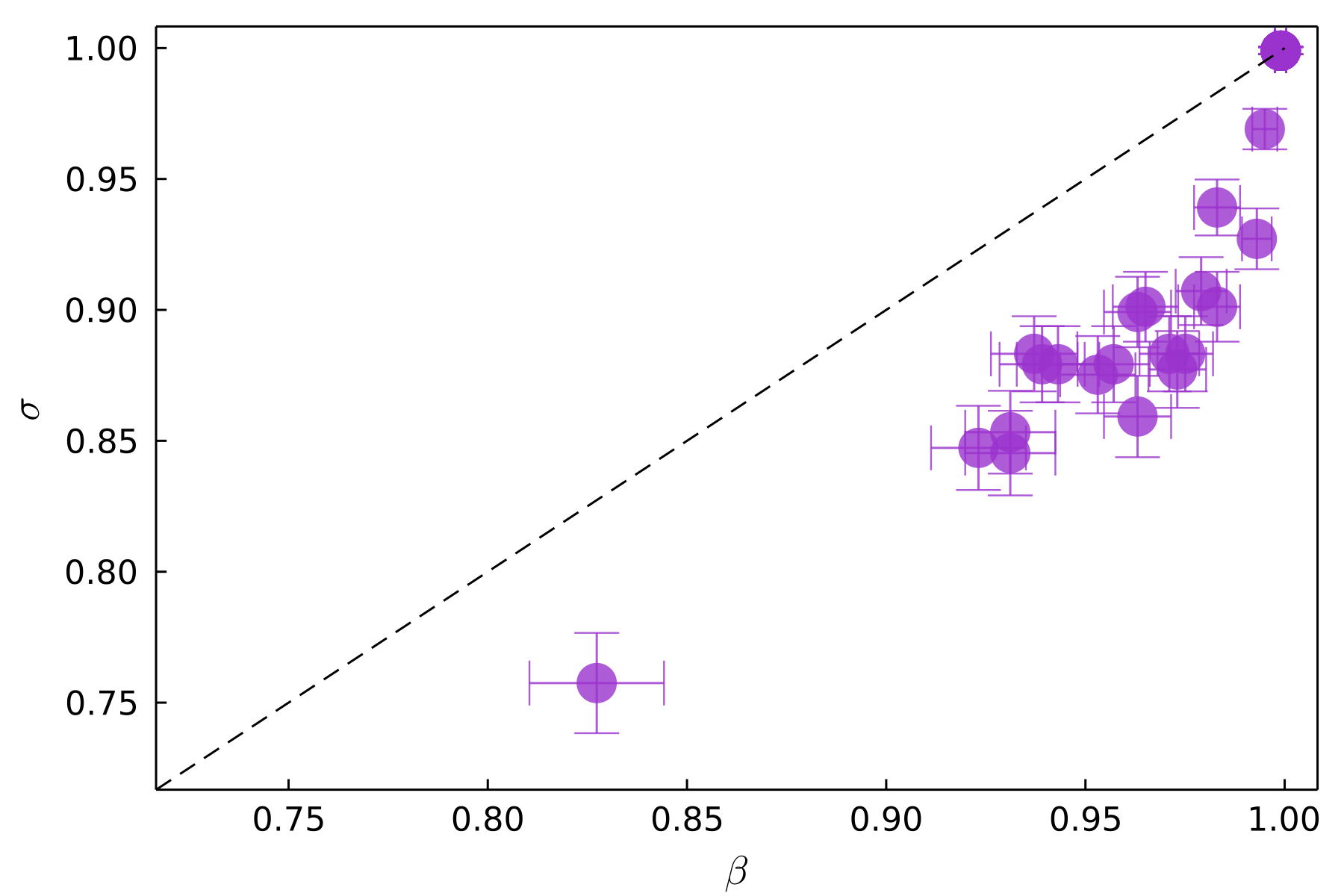}
       \caption{Comparison between the basin stability $\beta$ and the survivability $\sigma$ of all nodes in the IEEE 96 test case. It can be seen that the survivability is generally smaller or equal to the basin stability. The confidence intervals were calculate using ”add two successes and failures” \cite{add_two_successes_and_failures}.}
       \label{fig:snbs_surv_5}
    \end{figure}

\begin{figure}[H]
        \centering
            \begin{subfigure}[b]{0.49\textwidth}
                \includegraphics[width=\textwidth]{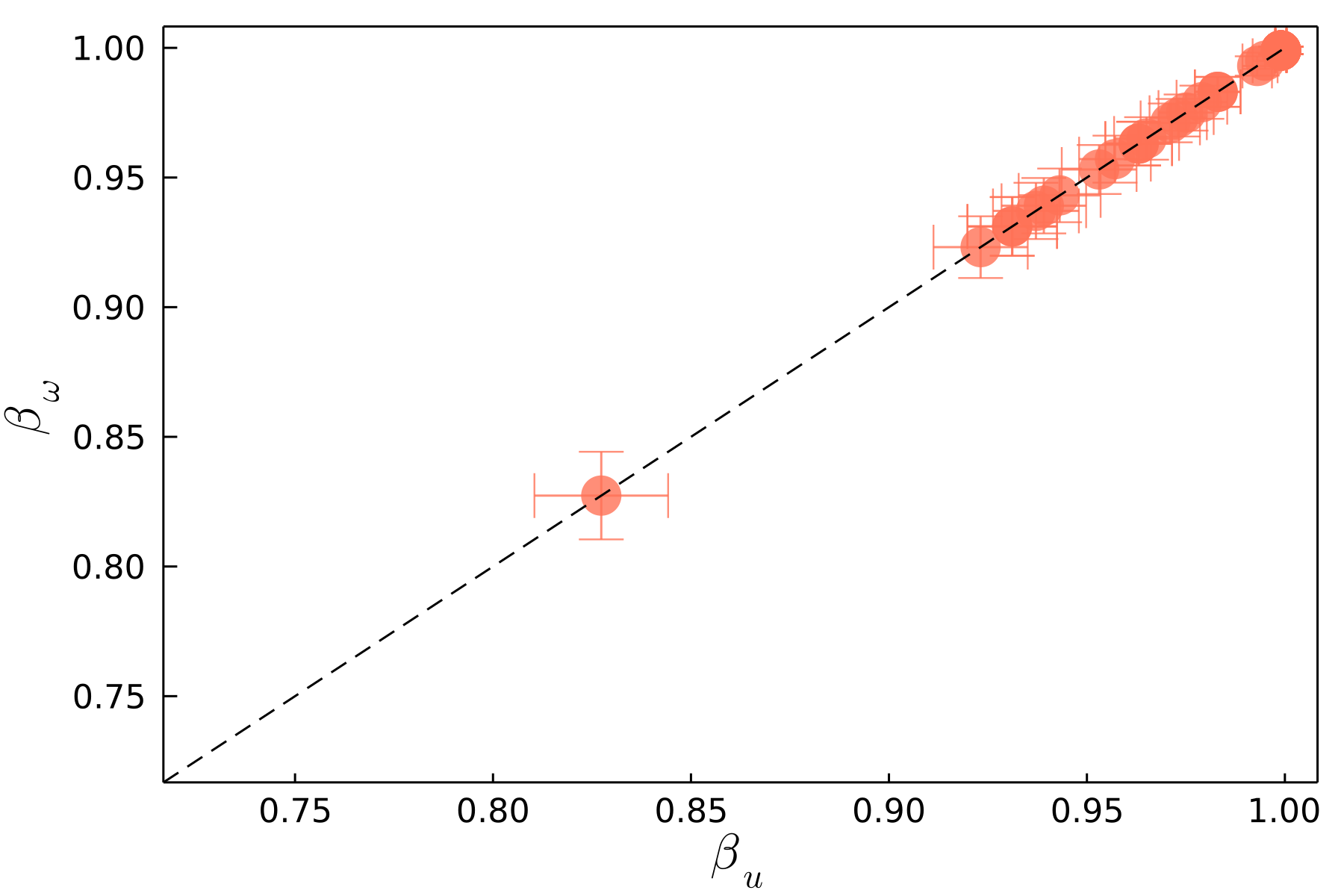}
                \caption{Basin Stability}
                \label{fig:96Bus_snbs_omgea_u_5}
            \end{subfigure}
            \hfill
            \begin{subfigure}[b]{0.49\textwidth}
                \includegraphics[width=\textwidth]{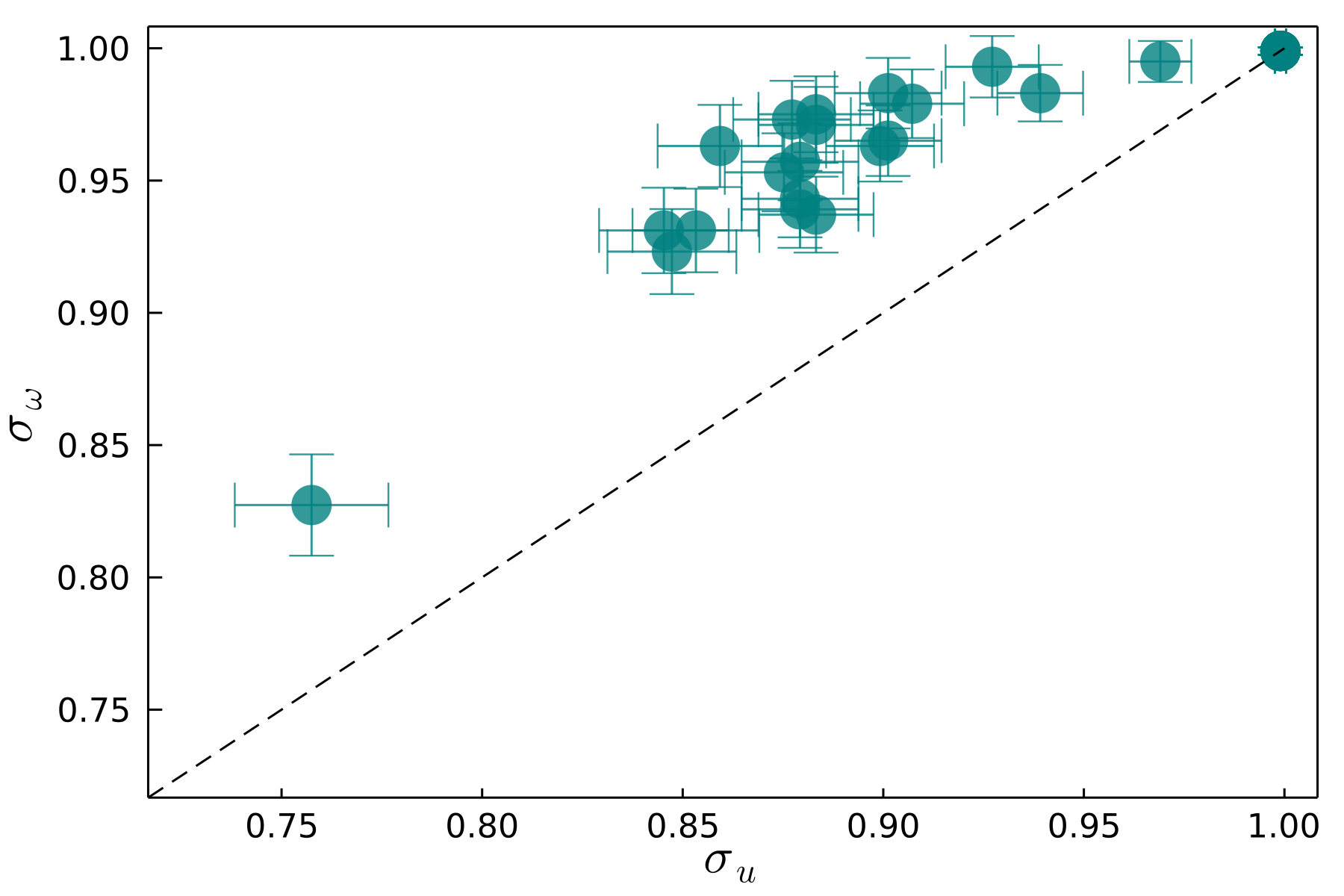}
                \caption{Survivability}
                \label{fig:96Bus_surv_omgea_u_5}
            \end{subfigure}
        \caption{Comparison between the basin stability and survivability of the IEEE 96 test case evaluated for the different conditions. Either only the voltage condition, denoted by an $u$ index, is evaluated or only frequency condition, denoted by the $\omega$ suffix. The right figure shows that the basin stability does not change when we evaluate different conditions, meaning that when the frequency does not return to the initial fixed point the voltage will also not and vice versa. The left figure shows that for the survivability evaluating only the frequency condition is not enough as there are many nodes in which voltage leaves the desired region while the frequency does not. The confidence intervals were calculate using ”add two successes and failures” \cite{add_two_successes_and_failures}.}
    \end{figure}

    To gain further information, we increase the strength of the frequency perturbations to $\Delta \omega = [-5,5]$, and the strength of the voltage perturbations to $\pm 25\%$ of the operation point level. The results of the basin stability and survivability are shown in the figures \ref{fig:hist_96Bus_snbs_5}-\ref{fig:96Bus_surv_omgea_u_5}. The histograms in the figures \ref{fig:hist_96Bus_snbs_5} and \ref{fig:hist_96Bus_surv_5} show that all PQ Node are globally stable meaning that they always return to the initial fixed point after a perturbation and never leave the desired region. For the normal form, some nodes also have basin stability and survivability of 1. But some of the differential nodes also become less stable.  Additionally, it was found that for all nodes the survivability $\sigma$ is smaller or equal to the basin stability $\beta$, which can be seen in Figure \ref{fig:snbs_surv_5} in the appendix.
    
    % Histograms
    
    It is interesting to check which of the EN 50160 Report requirements, for the voltage amplitude $v$ or the frequency $\omega$, were broken. It should be noted that the different $\beta$ and $\sigma$ results form the same set of initial perturbations, just the two different conditions were evaluated.
    
    We found that the basin stability remains the same when we evaluate the voltage or the frequency condition. This means that when the frequency does not return to the initial fixed point, the voltage will also not and vice versa. For the survivability, we found that only evaluating the frequency condition will overestimate the survivability. This indicates that there are many nodes whose voltages leave the desired region while the frequency does not. This evaluation is shown in the figures \ref{fig:96Bus_snbs_omgea_u_5} and \ref{fig:96Bus_surv_omgea_u_5} in the appendix.
    
%%%%%%%%%%%%%%%%%%%%%%%%%%%%%%%%%%%%%%%%%%%%%%%%%%%%%%%%%%%%%%%%%%%%%%%%%%%%%%%%%%%%%%%%%%%%%%    
    \subsection{Voltage Sweep}
\begin{figure}[H]
            \centering
            \begin{subfigure}[b]{0.49\textwidth}
                \includegraphics[width=\textwidth]{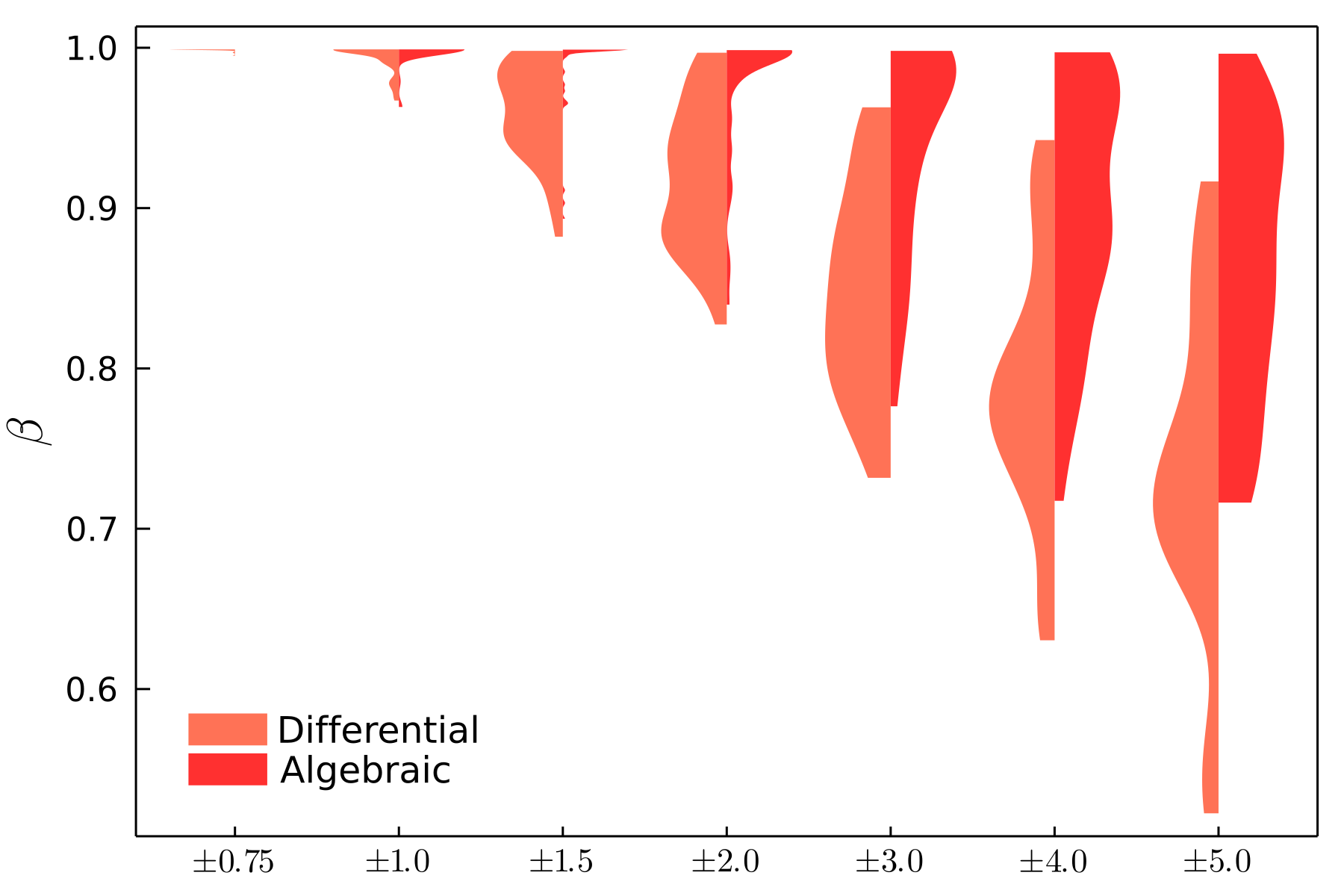}
                \caption{Basin Stability}
                \label{fig:96Bus_snbs_vol_vio}
            \end{subfigure}
            \hfill
            \begin{subfigure}[b]{0.49\textwidth}
                \includegraphics[width=\textwidth]{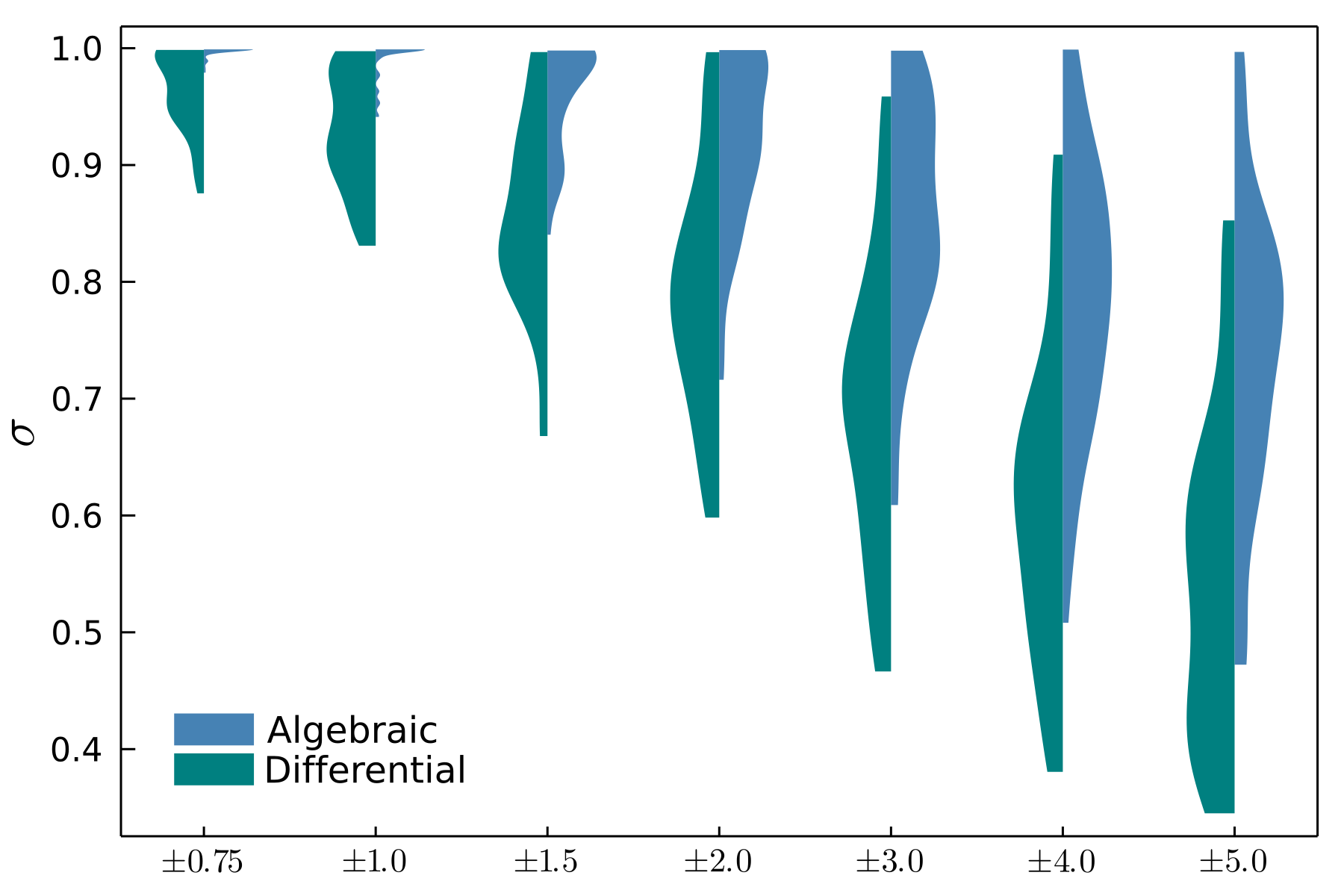}
                \caption{Survivability}
                \label{fig:96Bus_surv_vol_vio}
            \end{subfigure}
        \caption{Dependence of the basin stability and survivability on the strength of the voltage perturbations. The basin stability first starts to react to the perturbation with a level of $\pm 100 \%$ from the operation point. The basin stability is generally higher than the survivability. Differential nodes are less stable and survive less frequently than algebraic nodes.}
    \end{figure}

    As we have not yet seen an effect of voltage perturbations, we increase the maximal perturbations up to $\pm 500\%$ of the operation point level. Although only the voltage was perturbed, we still evaluate the frequency bounds for the survivability. The bounds are again chosen following the EN report, meaning that the threshold is $[-3, 2]$. The results of the voltage sweep are shown in the figures \ref{fig:96Bus_snbs_vol_vio} to \ref{fig:96Bus_surv_vol_vio}.
    
    Up to a perturbation level of $\pm 100 \%$, we do not see an effect on basin stability and survivability and all nodes are globally stable. Increasing the perturbation further, we see that the basin stability and survivability decrease. We discover that the dynamical nodes have a lower $\beta$ and $\sigma$ than the algebraic nodes. Just as before we see that the survivability $\sigma$ is smaller than the basin stability $\beta$.
    
    Although the voltage is crucial for the survivability of the IEEE96 test case, it is not the realistic voltage perturbations which destabilize them, as the voltage perturbations only start to play a role at perturbation strengths of around $\pm 100 \%$. This implies that the frequency perturbations, in the regime of the realistic perturbations, pushes the voltages out of the desired regime. This further illustrates that it is important to associate several stability classes with each other.
    
%%%%%%%%%%%%%%%%%%%%%%%%%%%%%%%%%%%%%%%%%%%%%%%%%%%%%%%%%%%%%%%%%%%%%%%%%%%%%%
    
\section{Conclusion}
    In this paper, we have introduced a novel method for probability distributions on manifolds that can generate random, initial conditions for probabilistic stability methods like basin stability and survivability. By introducing the new network measure, the spreadability, we have proven that the initial conditions are network local in a DAE sense and that the states are thus usable for single node basin stability and survivability. 
    
    We have utilized the method to calculate the basin stability and survivability of an ensemble of random synthetic power grids, which includes grid following as well as grid forming inverters. This was the first time basin stability and survivability were calculated for systems with constraints. We have applied the topological classification scheme, which was introduced by Nitzbon et. al. \cite{topology_algo}, and investigate if it still holds. We can report that, while the classes are still able to separate the nodes with respect to their stability properties, the degree is a much better predictor for the basin stability and especially the survivability. While the degree is a good predictor we want to draw attention to the fine structure inside of the distributions of the survivability which indicates that there might be another effect that affects it.
    
    Interestingly we were able to show the inverse effect of the degree on the survivability than the original paper \cite{topology_algo}. We found that the survivability of the nodes increases with the degree, whereas before $\sigma$ decreased with a higher degree. 
    
    Additionally, we do not see the frequent desynchronization events which are normally observed when only the swing equation is used as the nodal model. An important destabilizing mechanism is that the systems frequently run into infeasible power flow solutions, meaning that the constraints can no longer be fulfilled. This could potentially be avoided by adding fast energy storage to the loads and grid following inverters to reach power balance. A more detailed model of loads is needed for such an investigation.
    
    We also found out that algebraic nodes are more stable and survive more often after a perturbation than the differential nodes. But they still play a crucial role in the destabilization of the system as they constraint the power flow and thus create infeasible solutions. Generally, the basin stability is higher than the survivability.
    
    Finally, we have applied the method to calculate the basin stability of the IEEE 96 system. For smaller, more realistic perturbations we see that again the algebraic nodes are more stable than the differential nodes. In fact that all algebraic nodes have basin stability and survivability of 1. Thus we have increased the strength of the voltage perturbation and found that voltage perturbations only start to play a role if we perturb the system from its fixed point $\pm 100 \%$, meaning that realistic voltage perturbations do not play a role in the destabilization of the system. 
    
\section*{Code availability}
    All code to reproduce the results and figures of this paper is available at the GitHub repository \url{https://github.com/PIK-ICoNe/AmbientForcing_Paper_Companion} or on zenodo \url{https://doi.org/10.5281/zenodo.5901260}. The implementation of the Ambient Forcing Method can be found in a separate GitHub repository \url{https://github.com/PIK-ICoNe/AmbientForcing.jl}.
    
\section*{Acknowledgments}
    A. Büttner acknowledges support by the German Academic Scholarship Foundation.
    All authors gratefully acknowledge the European Regional Development Fund (ERDF), the German Federal Ministry of Education and Research and the Land Brandenburg for supporting this project by providing resources on the high-performance computer system at the Potsdam Institute for Climate Impact Research.
    
%%%%%%%%%%%%%%%%%%%%%%%%%%%%%%%%%%%%%%%%%%%%%%%%%%%%%%%%%%%%%%%%%%%%%%%%%%%%%%%%%%%%%%%%%%%%    
    
\printbibliography[heading=bibintoc, title = {Bibliography}]
 \newpage
\section*{Appendix}
    \subsection*{Phase Amplitude Oscillators Parameters}
    \label{sec:parameter}
    \begin{table}[H]
    \centering
        \begin{tabular}{|l|l|l|l|l|l|}
            \hline
                      & $A$             & $B$  & $C$      & $G$  & $H$       \\ \hline
            $u$       & 0.075 $\cdot Q$ & 1$i$   & -0.0625  & 0    & -0.0125 \\ \hline
            $\omega$  & 5 $\cdot P$     & -1   & 0        & -5   & 0       \\ \hline
        \end{tabular}
        \caption{Normal form coefficients in terms of the original parameters. $A$ depends on the active and reactive power sets points, $P$ and $Q$ respectively, at the node node.}
        \label{tab:parameters}
    \end{table}   
%%%%%%%%%%%%%%%%%%%%%%%%%%%%%%%%%%%%%%%%%%%%%%%%%%%%%%%%%%%%%%%%%%%%%%%%%%%%%%%%%%%%%%%%%%%%%%     
    \subsection*{IEEE96} 
    
%%%%%%%%%%%%%%%%%%%%%%%%%%%%%%%%%%%%%%%%%%%%%%%%%%%%%%%%%%%%%%%%%%%%%%%%%%%%5    
    
    \subsection*{Amplification}
    \label{sec:amp}
    \begin{figure}[H]
            \centering
            \begin{subfigure}[b]{0.49\textwidth}
                \includegraphics[width=\textwidth]{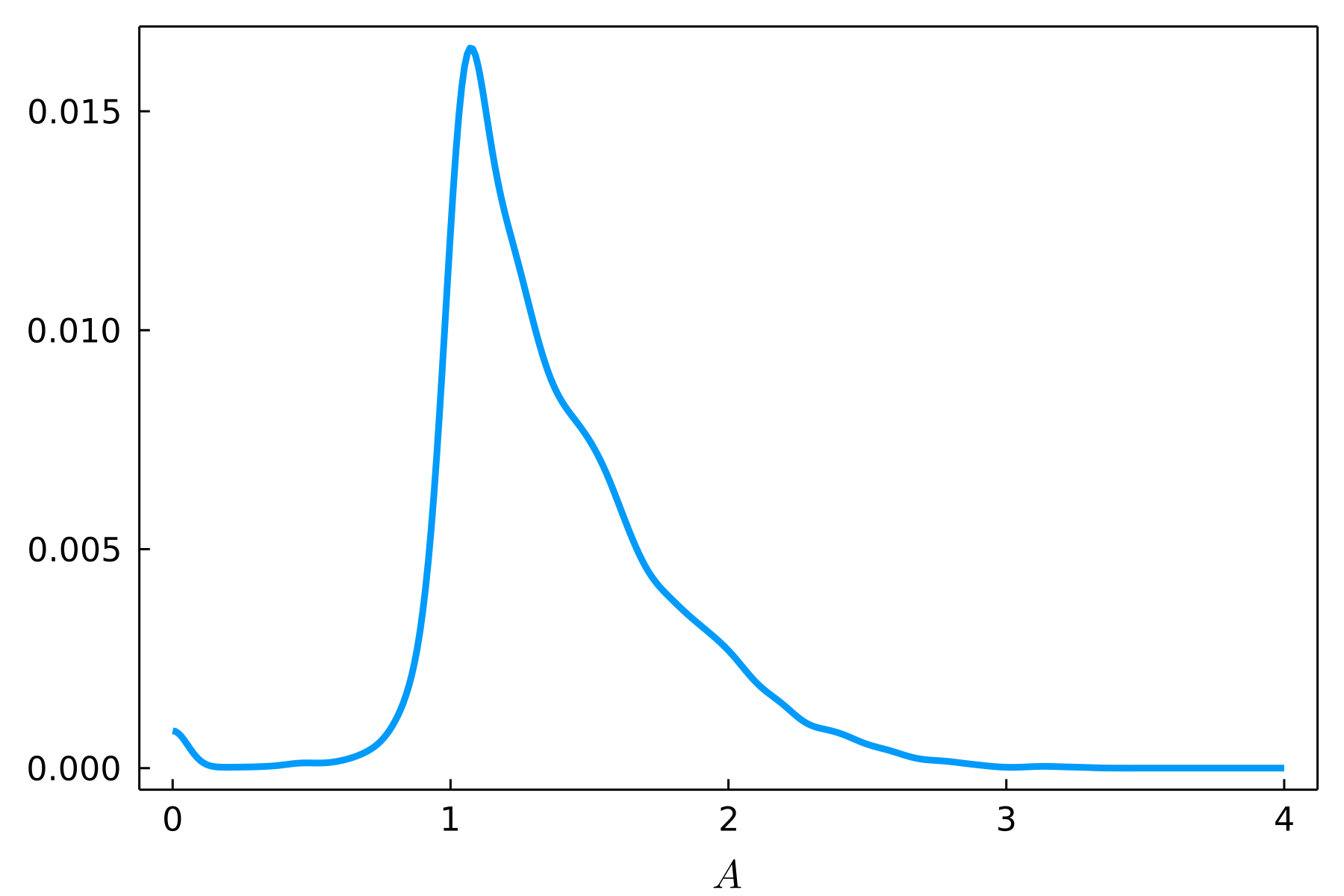}
                \caption{Density Estimation}
            \end{subfigure}
            \hfill
            \begin{subfigure}[b]{0.49\textwidth}
                \includegraphics[width=\textwidth]{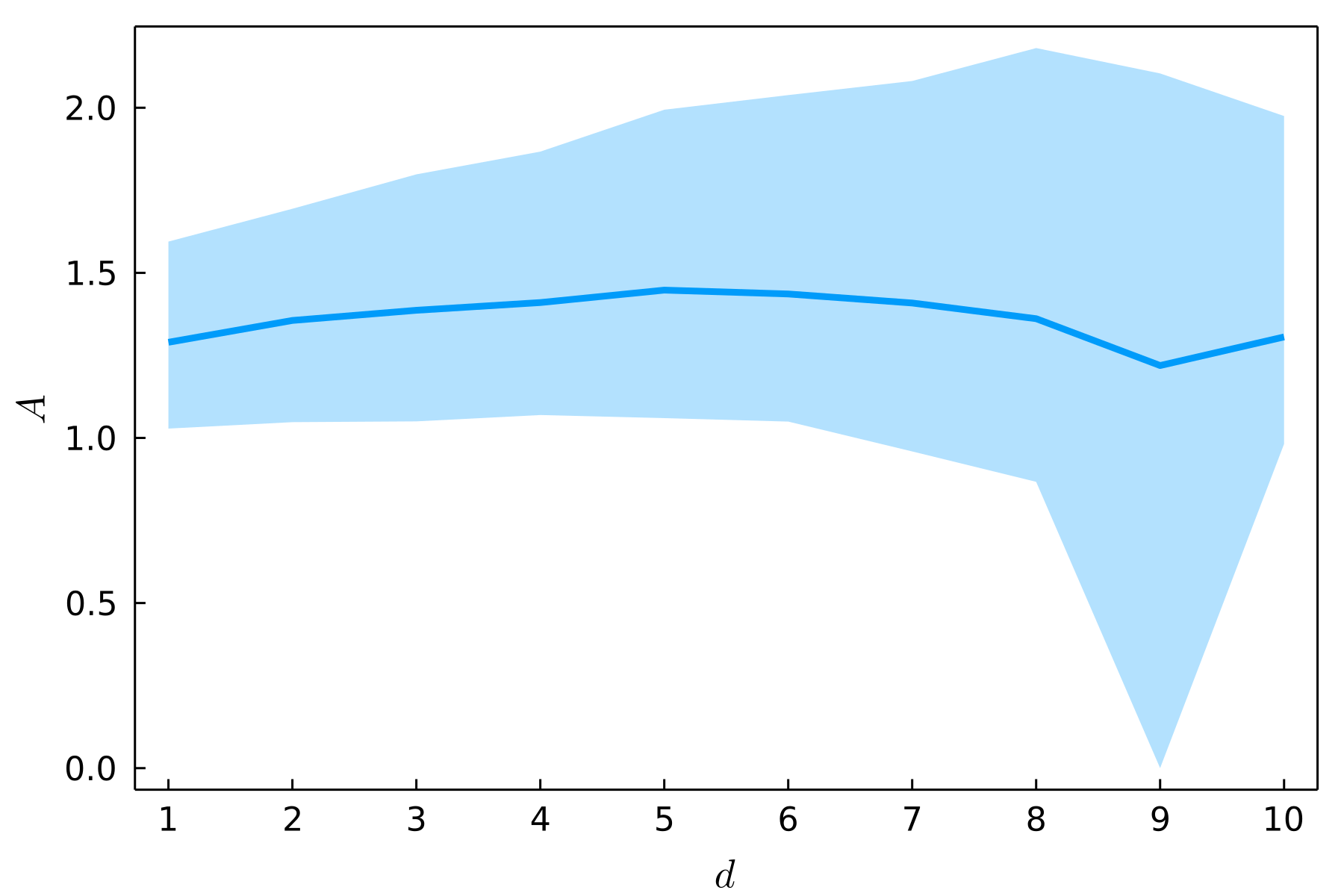}
                \caption{Degree Dependence}
            \end{subfigure}
        \caption{Distribution of the amplification $A$. The mode of the distributions lies at 1.1. The second smaller peak at $A=0$ occurs because of the slack nodes whose voltage can not be perturbed. In the right figure, we can see that the mean amplification slightly increases with the degree. For higher degrees, we can see that the size of percentiles increases. This dependency on the degree can not explain the strong increase of the survivability with the degree. If this would have been the case we would need to see a decrease of $A$ with the degree.}
        \label{fig:amp}
    \end{figure}

    \subsection*{Power Grids and Example Trajectories}
    In this section, we will show some exemplary trajectories which occur after perturbations on single nodes, as well as the corresponding synthetic power grids. The nodes in the power grid are classified using the algorithm from \cite{topology_algo}. Algebraic nodes are depicted as triangles while differential nodes are shown as circles. The size of the nodes corresponds to the single node basin stability.

    \begin{figure}[H]
            \centering
             \begin{subfigure}[b]{0.49\textwidth}
                \includegraphics[width=\textwidth]{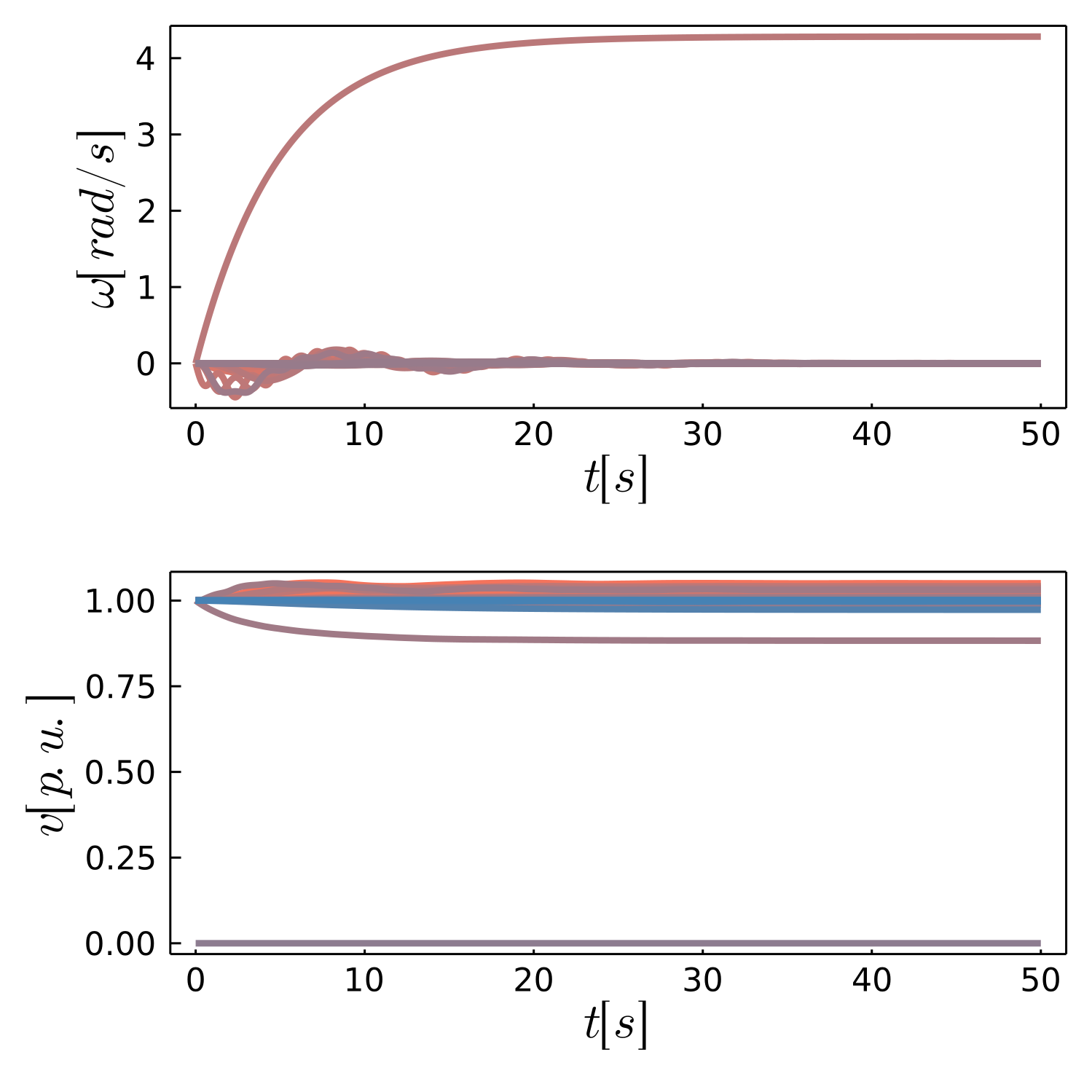}
                \caption{Perturbation on a Sparse Sprout}
            \end{subfigure}
            \hfill
            \begin{subfigure}[b]{0.49\textwidth}
                \includegraphics[width=\textwidth]{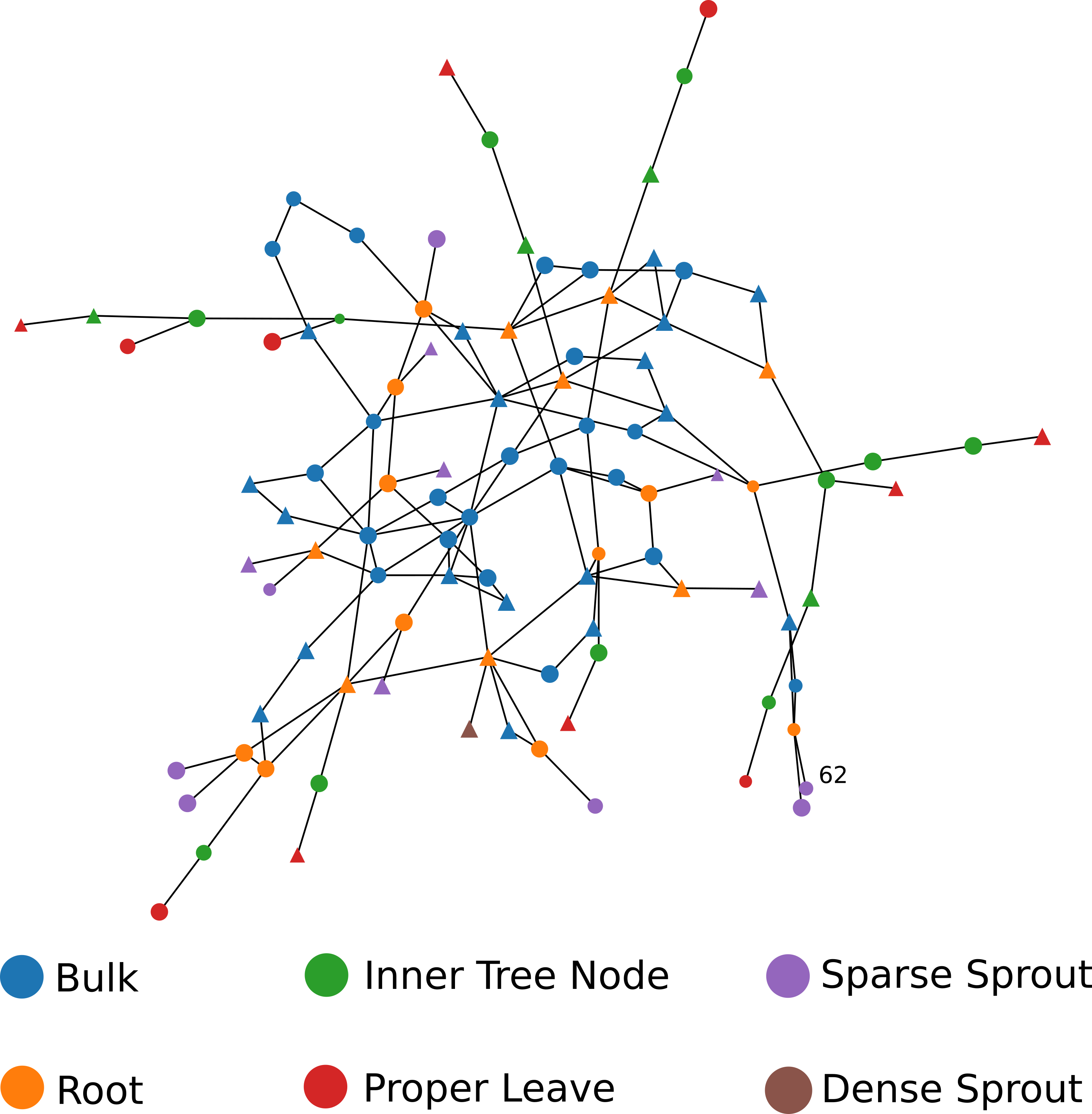}
                \caption{Power Grid}
            \end{subfigure}
        \caption{Desynchronization and a short circuit after a perturbation on the sparse sprout 62.}
        \label{fig:desync_vol_drop}
    \end{figure} 
    
    \begin{figure}[H]
            \centering
            \begin{subfigure}[b]{0.49\textwidth}
                \includegraphics[width=\textwidth]{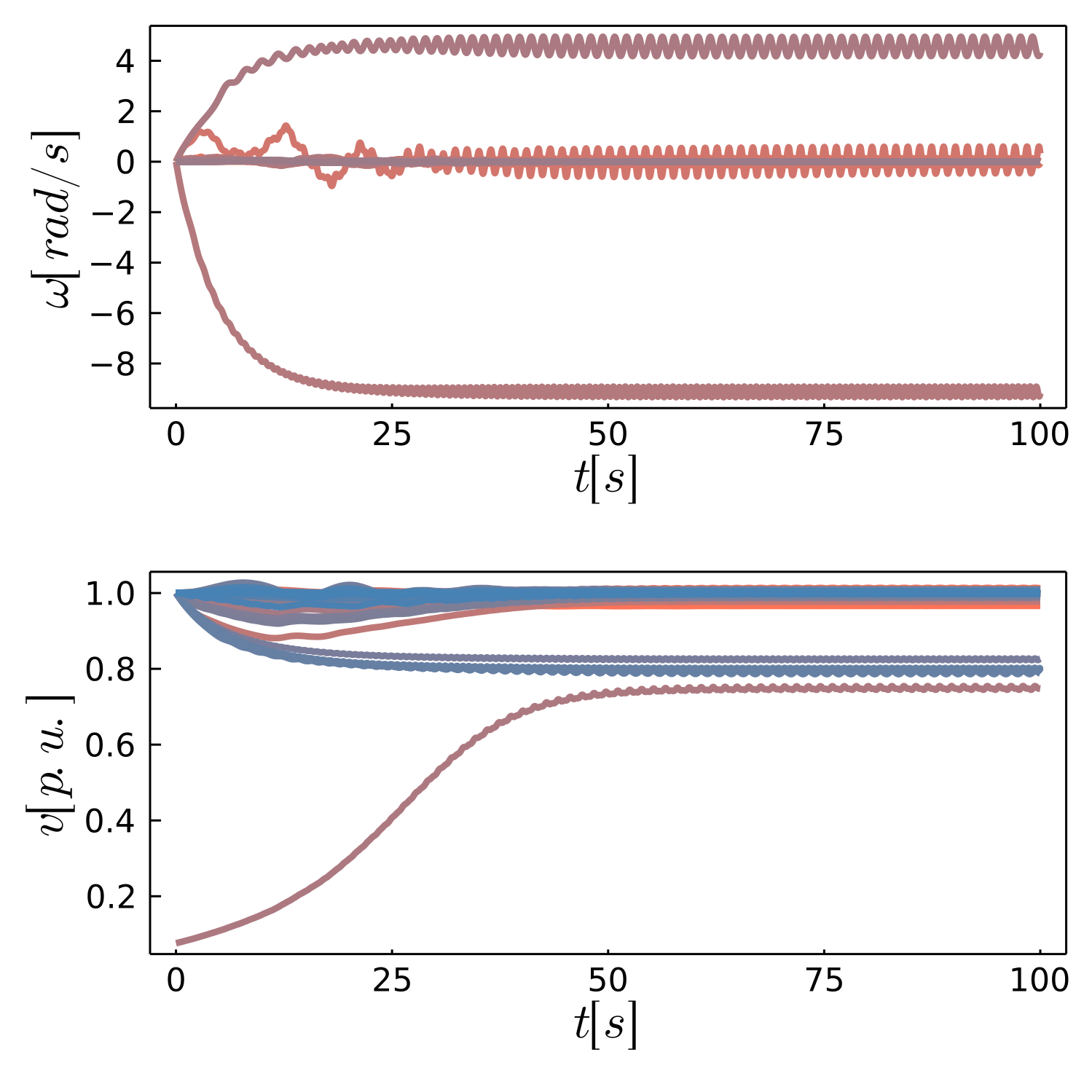}
                \caption{Perturbation on an Inner Tree Node}
            \end{subfigure}
            \hfill
            \begin{subfigure}[b]{0.49\textwidth}
                \includegraphics[width=\textwidth]{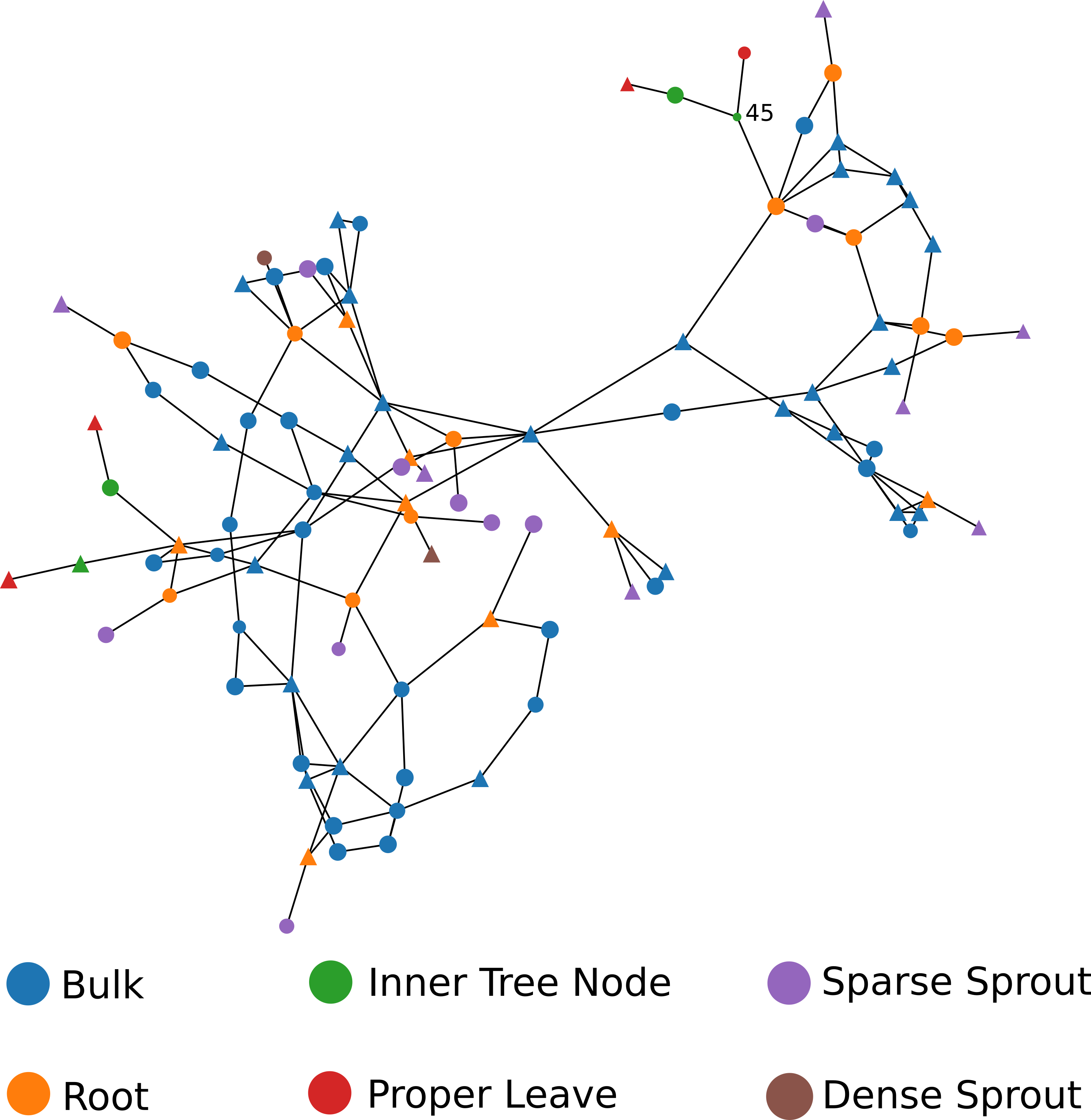}
                \caption{Power grid}
            \end{subfigure}
        \caption{Perturbation on an inner tree node that leads to a desynchronization and voltage drops.}
        \label{fig:desync_vol_drop_double}
    \end{figure}

\end{document}